%% file: main.tex
\documentclass[%
 aip,
 amsmath,amssymb,
reprint,%
]{revtex4-1}

\usepackage{graphicx}
\usepackage{dcolumn}
\usepackage{bm}

\usepackage[utf8]{inputenc}
\usepackage[T1]{fontenc}
\usepackage{mathptmx}
\usepackage{etoolbox}

\makeatletter
\def\@email#1#2{%
 \endgroup
 \patchcmd{\titleblock@produce}
  {\frontmatter@RRAPformat}
  {\frontmatter@RRAPformat{\produce@RRAP{*#1\href{mailto:#2}{#2}}}\frontmatter@RRAPformat}
  {}{}
}%
\makeatother
\begin{document}

\preprint{AIP/123-QED}

\title[Reconstruction of 4D Mitral Regurgitation Hemodynamics from Sparse Planar Data using DeepONets with TTA]{Reconstruction of 4D Mitral Regurgitation Hemodynamics from Sparse Planar Data using Deep Operator Networks with Test-Time Adaptation}
\author{Jakob Marcel Hoffmann}
\affiliation{ 
Institute of Fluid Mechanics, Karlsruhe Institute of Technology, 76131 Karlsruhe, Germany
}%
\author{Yosuke Hasegawa}
\affiliation{%
Center for Research on Innovative Simulation Software, Institute of Industrial Science, The University of Tokyo, Japan
}%

\author{Alexander Stroh}
\affiliation{ 
Institute of Fluid Mechanics, Karlsruhe Institute of Technology, 76131 Karlsruhe, Germany
}%

\date{\today}

\begin{abstract}
\input{sections/abstract}
\end{abstract}

\maketitle

\section{Introduction}
\input{sections/intro}

\section{Methodology}
\input{sections/methods}

\section{Results}
\input{sections/results}

\section{Conclusion}
\input{sections/conclusion}

\begin{acknowledgments}
This work was supported by the JSPS invitational fellowship for researchers in Japan, ID S25051, the UTokyo-IIS internship support program, and the DAAD-PROMOS scholarship.
\end{acknowledgments}

\section*{Data Availability Statement}

The data that support the findings of this study are available from the corresponding author upon reasonable request.

\section*{Author Contributions}
\textbf{Jakob Marcel Hoffmann}: Methodology, Software, Validation, Formal analysis, Investigation, Data Curation, Writing – original draft, Visualization 

\textbf{Yosuke Hasegawa}: Conceptualization, Methodology, Resources, Writing – review \& editing, Supervision, Funding acquisition

\textbf{Alexander Stroh}: Conceptualization, Methodology, Resources, Writing – original draft, Supervision, Funding acquisition

\section*{Conflict of Interest}
The authors have no conflicts to disclose.

\appendix
\section*{Appendix}
\subsection{\label{sec:meshdetails} CFD Mesh Details}
\input{sections/simdetails}

\subsection{\label{sec:validation} CFD Validation}
\input{sections/validation}

\subsection{Error Metrics Definitions}
\input{sections/defs}

\subsection{\label{sec:appresults} Additional Results}
\input{sections/appresults}

\clearpage

\bibliography{references}

\end{document}

%% file: sections/abstract.tex
Quantifying mitral regurgitation severity remains limited by the assumptions
of clinical flow convergence methods, while high-fidelity simulation and
volumetric velocimetry are too slow for routine use. We investigate whether a
learned solution operator can reconstruct transient three-dimensional
transvalvular hemodynamics from the sparse observation an in-vitro experiment
actually provides: a single planar velocity slice and two boundary pressure
traces. A Deep Operator Network is pretrained on an experimentally benchmarked
URANS database spanning eleven mitral regurgitation orifice phantoms, learning
a mapping from a masked two-component planar velocity snapshot to the
surrounding volumetric field, and is subsequently adapted to unseen target
cases by fine-tuning on their sparse measurements. Adaptation reliably corrects
the flow topology within the supervised plane, reorienting a strongly eccentric
jet that the pretrained operator predicts as straight, and yields full-field
predictions in minutes rather than the days required by the underlying
simulations. Its influence decays sharply with distance from that plane,
however: measured against phase-resolved particle image velocimetry, the
reconstruction error rises from 24.6\% at $2\, \mathrm{mm}$ to 52.6\% at
$6\, \mathrm{mm}$, and the resulting mismatch between corrected and
uncorrected layers degrades physical consistency. Single-plane supervision thus
constrains the observed plane far more effectively than the surrounding volume,
which we identify as the principal obstacle to coherent 4D reconstruction from
sparse planar data.

%% file: sections/intro.tex
Mitral regurgitation (MR) is a progressive cardiovascular condition characterized by inadequate mitral valve (MV) closure, where accurate quantification of disease severity is crucial for clinical decision-making.\cite{12} Clinically, severity is most often graded with the flow convergence method, which infers the regurgitant volume from a proximal isovelocity surface area (PISA) assumed to be hemispherical. The underlying idealization of a small, circular orifice with a radially symmetric approach flow deteriorates with increasing orifice area and non-circular shape, and its evaluation additionally depends on the operator-selected aliasing velocity; jointly, these effects have been shown to underestimate the regurgitant volume by up to 52\% in a pulsatile in-vitro reference experiment, sufficient to misclassify moderate regurgitation as mild.\cite{Leister2025} Highly controlled in-vitro setups utilizing Mitral Regurgitation Orifice Phantoms (MROPs) have therefore been developed to investigate regurgitant jets under reproducible conditions.\cite{28,Leister2025} By varying orifice shape and size, these phantoms permit controlled investigation of the relationship between defect geometry, jet topology, and regurgitation severity.

To complement these physical models, computational hemodynamics has rapidly evolved, progressing from simplified CFD validated against in-vitro measurements\cite{Quaini2012,Sonntag2014,Leister2025} to image-based approaches resolving whole-heart turbulence from multi-series
cine-MRI.\cite{Jamil2017,Collia2019,Bennati2023b,Bennati2024} State-of-the-art fluid-structure interaction (FSI) models further leverage subject-specific anatomies to investigate chordal mechanics and computationally evaluate transcatheter repairs.\cite{Toma2017,Caballero2018,Caballero2020,Dabiri2022}

Despite these advances, the analysis of complex MV flow dynamics remains subject to severe experimental and computational bottlenecks. Although two-component 2D particle image velocimetry (2D2C PIV) provides high-resolution planar velocity fields, reconstructing a flow volume requires sequential acquisition of many planes, remains restricted to optically accessible regions, and yields neither the out-of-plane velocity
component nor the pressure field. High-fidelity transient CFD and FSI simulations overcome these observational limits but demand high spatiotemporal resolution\cite{Bennati2023b}, computationally intensive coupled solvers\cite{Toma2017}, and accurately prescribed boundary conditions that complicate automated workflows.\cite{Caballero2020} Since a new simulation is generally required for every geometry and operating condition, and each can require days or weeks of computation, clinical applications are driven toward reduced-order or steady-state approximations.\cite{Vellguth2022,Kirchner2025}

Machine-learning-based flow reconstruction offers a route to bridge sparse measurements and expensive fully resolved simulations.\cite{56} Sparse-sensor reconstruction targets precisely this setting, estimating a high-dimensional field from few, noise-corrupted measurements by exploiting the compressibility of the underlying dynamics.\cite{manohar2018data,karnik2024constrained} Classical formulations project onto a linear basis obtained by Proper Orthogonal Decomposition (POD) or Dynamic Mode Decomposition and recover the modal coefficients from a least-squares fit at the sensor locations, as in Gappy POD.\cite{everson1995karhunen,willcox2006unsteady} Because linear
reconstructions are limited for strongly nonlinear flows\cite{bucquet2025sparse}, extensions such as Bayesian Gappy POD\cite{bertram2024fusing}, shallow decoders mapping sensor readings directly to the full field\cite{erichson2020shallow}, and Gappy autoencoders\cite{kim2024gappy} have been introduced. These methods nevertheless reconstruct within a single fixed configuration, as a change of geometry generally requires a new basis or a retrained decoder. Physics-Informed Neural Networks (PINNs) instead regularize reconstruction using the governing equations and have been applied to inverse flow problems\cite{50}, including three-dimensional cardiac velocity and pressure fields recovered from sparse, noisy Doppler observations.\cite{Wong2025} They remain instance-specific, however, requiring reoptimization for each new configuration.

Operator-learning approaches address this limitation by learning mappings between families of input and solution functions rather than approximating an individual solution. Deep Operator Networks (DeepONets) employ separate branch and trunk networks to learn such nonlinear solution operators and permit rapid inference once trained.\cite{37} Their use in cardiovascular hemodynamics has recently expanded to parameterized arterial flow\cite{Hong2026}, three-dimensional velocity and pressure prediction in idealized aortic aneurysms\cite{CruzGonzalez2026}, and multifidelity formulations combining simulations with limited in-vitro data in stenosed vessels.\cite{Velikorodny2025} These studies predominantly formulate a forward surrogate problem, mapping physiological parameters or boundary conditions to the resulting hemodynamic field. The problem considered here is instead an inverse reconstruction problem, in which a spatially localized and incomplete observation must be used to infer the surrounding three-dimensional transient field.

A remaining challenge of operator learning is its dependence on the training distribution: prediction error can increase substantially for out-of-distribution (OOD) inputs\cite{18}, which is particularly relevant when operators pretrained on simulations are applied to experiments subject to measurement noise, temporal sparsity, and geometric discrepancies. Test-Time Adaptation (TTA) addresses this by adapting a pretrained operator to the specific target instance rather than training from scratch, exploiting governing equations, sparse target observations, or both.\cite{Goswami2022Transfer,33,66} In previously reported applications, however, the adapting observations are distributed across the domain, whereas the planar velocimetry considered here constrains a single slice of an otherwise unobserved volume.\cite{66} The extent to which such a spatially confined constraint propagates into the surrounding three-dimensional field has, to the authors' knowledge, not been assessed.

Building on this foundation, we develop a DeepONet framework for reconstructing transient three-dimensional hemodynamic fields from spatially sparse observations representative of an in-vitro MR experiment. The operator is pretrained exclusively on an experimentally benchmarked URANS database spanning multiple MROP geometries, learning a generalized mapping from localized 2D2C velocity observations to the surrounding volumetric field. It is subsequently driven by, and fine-tuned on, measurements from the physical simulator, so that its application constitutes a simulation-to-experiment transfer subject to measurement noise, temporal sparsity, and the geometric discrepancy between the rigid CFD orifice plates and the flexible physical foil.


%% file: sections/methods.tex
\subsection{CFD-based Synthetic Dataset}

\subsubsection{Geometry and Computational Domain}

\begin{figure*}
\centering
    \includegraphics[width=1\textwidth]{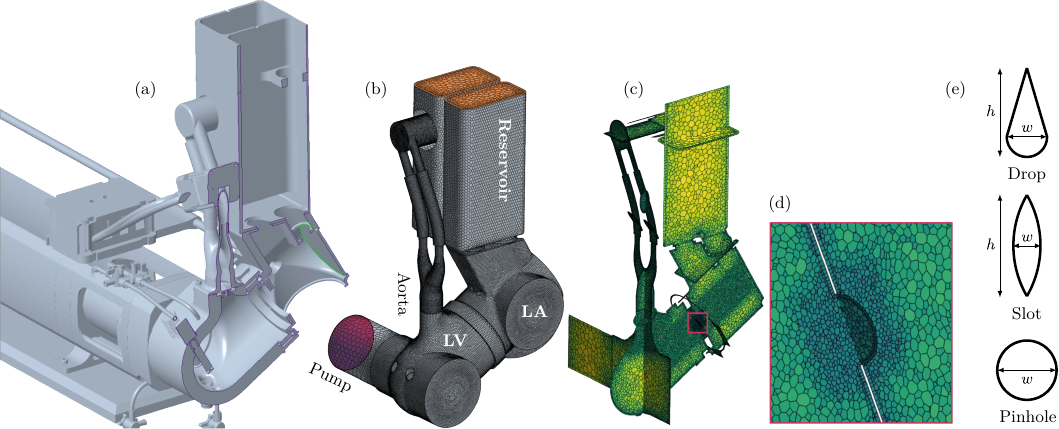}
    \caption[Hemodynamic simulator geometry]{\label{fig:mesh} Hemodynamic simulator geometry. (a) CAD cutaway of the chambers, (b) meshed internal fluid domain, (c) mesh slices, (d) magnified Pinhole-L orifice mesh, (e) MROP shapes with height and width markers. LV = Left Ventricle, LA = Left Atrium.}
\end{figure*}

The proposed framework relies on a training database of simulated hemodynamics, designed to mirror the physical in-vitro hemodynamic simulator as shown in Fig.~\ref{fig:mesh},a. The computational fluid domain was extracted from the computer-aided design (CAD) model of the experimental in-vitro setup and simplified in a few non-critical locations. At the core of the simulator is the Mitral Regurgitation Orifice Phantom (MROP), positioned between the left ventricle (LV) and left atrium (LA). The MROPs are manufactured from a $0.5\,\mathrm{mm}$ thin PVC film. Their purpose is to reproducibly replicate the insufficient closure of the mitral valve during systole. Because the MROPs simulate a fixed, continuously leaky state rather than a dynamic valve, the experimental setup does not feature an aortic valve to allow the LV to more easily refill during diastole. In reality, the thin polymer film is flexible and deforms into a spherical dome under the transvalvular pressure gradient. 
This effect was neglected in the present CFD model, in which the MROP was represented as a fixed, rigid wall.

\begin{figure}
    \centering
    \includegraphics[width=1\linewidth]{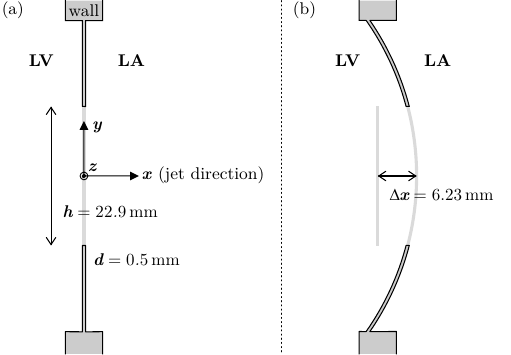}
    \caption{\label{fig:slotlbent} Side-view of rigid orifice plate geometries used in CFD. (a) Height $h$ and plate wall thickness $d$ for Slot-L. The origin of Cartesian coordinates is defined as this center of the flat opening for all MROPs. (b) Offset $\Delta x$ caused by spherical plate wall of Slot-L-Bent. }
\end{figure}

\begin{table}
\caption{\label{tab:mrops} Mitral Regurgitation Orifice Phantom (MROP) sizes. S, M, L, and XL designate Small, Medium, Large, and Extra-Large. Modified to include Drop-XL and EccJet. \cite{Leister2025} For the eccentric case, these values define the ellipse formed on the flat orifice plate.}
\begin{ruledtabular}
\begin{tabular}{lrrr}
Shape & height $h$ in mm & width $w$ in mm & area in mm$^2$ \\
\hline
Drop-S & 9.7 & 4.3 & 27.1 \\
Drop-M & 13.7 & 6.5 & 52.1 \\
Drop-L & 19.8 & 9.0 & 108.4 \\
Drop-XL & 27.6 & 13.0 & 215.5 \\
Slot-S & 11.1 & 3.3 & 27.0 \\
Slot-M & 14.0 & 4.5 & 44.8 \\
Slot-L(-Bent) & 22.9 & 7.3 & 115.1 \\
Pinhole-S & - & 4.7 & 17.1 \\
Pinhole-M & - & 8.7 & 58.8 \\
Pinhole-L & - & 12.2 & 116.7 \\
EccJet & 10.0 & 8.0 & 62.8 \\
\end{tabular}
\end{ruledtabular}
\end{table}

To establish a comprehensive ``prior knowledge'' distribution of transvalvular hemodynamics, a dataset comprising 11 distinct MROP geometries similar to the ones proposed by Leister~\textit{et al.}~\cite{Leister2025} was generated. Simulating this diverse array of conditions ultimately enabled the neural network to infer flow in unobserved spatial regions. The geometries spanned three primary classes (Drop, Slot, and Pinhole) as shown in Fig.~\ref{fig:mesh},e across varying severities (Small, Medium, Large, and Extra-Large) as defined in Table~\ref{tab:mrops}. To test the extrapolation capabilities of the machine learning framework, three cases were excluded from the training set: the oversized Drop-XL and the Slot-L-Bent for validation, and the EccJet for later evaluation during the application phase. The Slot-L-Bent incorporates a predefined spherical wall profile mirroring the maximum physical deformation of the foil during systole. The orifice plate geometry of Slot-L-Bent is compared against Slot-L in Fig.~\ref{fig:slotlbent}. Additionally, an eccentric case (EccJet) was included due to availability of stacked 2D2C planes (3D2C) phase-resolved measurement data. Unlike the thin films, the EccJet MROP is a $12\,\mathrm{mm}$ thick, 3D-printed plate featuring an $8\,\mathrm{mm}$ cylindrical through-hole tilted at a $37.5^\circ$ angle relative to the surface normal.

The spatial domain for the MROPs was discretized as a polyhedral mesh generally containing roughly 400,000 cells (Fig.~\ref{fig:mesh}, b,c,d). The Pinhole-L configuration utilized a denser mesh comprising approximately 1,000,000 cells, as the finer grid generated during the mesh independence study was retained for the final dataset. The meshes featured a single wall-adjacent prism layer and various regional refinements to optimize the trade-off between spatial resolution and computational cost. More detailed information regarding mesh settings, mesh independence, and near-wall resolution are provided in Appendix \ref{sec:meshdetails}

\subsubsection{Solver Configuration}

The numerical simulations were conducted using the commercial finite volume solver Simcenter STAR-CCM+ (Version 2406). The working fluid was prescribed to match the physical $30\,\mathrm{vol}\%$ glycerol and $70\,\mathrm{vol}\%$ water mixture used in the simulator.~\cite{Leister2025} It was modeled as an incompressible Newtonian fluid with a constant density of $\rho=1086.0\,\mathrm{kg/m^3}$ and a dynamic viscosity of $\mu = 2.9961\cdot 10^{-3}\,\mathrm{Pa\cdot s}$ at $20\,^\circ\mathrm{C}$.~\cite{59} 

For the generation of the CFD dataset, a fully resolved DNS in space and time is computationally prohibitive due to the small cell sizes required to resolve the thin orifice geometry and the correspondingly small time steps imposed by the CFL constraint at high jet velocities. Instead, the fluid flow was modeled by the Unsteady Reynolds-Averaged Navier-Stokes (URANS) equations. Through the Reynolds-decomposition, this approach separates the flow into resolved mean fields (e.g., $\overline{p}(x,t)$) and modeled turbulent fluctuations, which were closed using the Menter $k$-$\omega$-SST two-equation turbulence model.~\cite{42} Because the simulated cardiac flow is periodic, this Reynolds-averaging effectively yields a phase-averaged representation of the cardiac cycle. For notational simplicity, the overline $\overline{\square}$ designating these averaged quantities is omitted throughout this work. 
To provide appropriate boundary conditions for the momentum and turbulence equations at the walls, the ``All-$y^+$'' blended wall treatment was utilized.~\cite{siemens_digital_industries_software_simcenter_2024} The transient nature of the flow inside the chamber geometry results in varying near-wall velocities across the domain and over the cardiac cycle, leading to a broad range of non-dimensional wall distances ($y^+$) as detailed in Appendix~\ref{sec:meshdetails}. The ``All-$y^+$'' approach addresses this spatial and temporal variation by applying continuous blended wall functions that dynamically adapt to the local $y^+$ value. This ensures a consistent calculation of wall shear stress and turbulence quantities regardless of whether a given near-wall cell falls within the viscous sublayer, the buffer region, or the logarithmic layer.

The governing equations were solved sequentially utilizing a second-order segregated flow solver equipped with a SIMPLE algorithm. Conservative under-relaxation factors of 0.8 for velocity and 0.2 for pressure were employed. Temporal discretization relied on an implicit unsteady, second-order scheme. Regarding time-stepping, the solver advanced to the subsequent time-step once an asymptotic convergence criterion based on the stabilization of the mass flow through the orifice was satisfied. The time-step size was adaptive, maintaining a maximum CFL number of 0.5. In the case of Slot-L as an example, these settings resulted in 134,513 time-steps and 7,546,330 solver iterations over three cardiac cycles ($2.25\,\mathrm{s}$). In general, computing a single cardiac cycle demanded approximately 60 hours of wall-clock time on 64 cores (AMD EPYC 7662). 

\subsubsection{Boundary and Initial Conditions}

\begin{figure}
    \centering
    \includegraphics[width=1\linewidth]{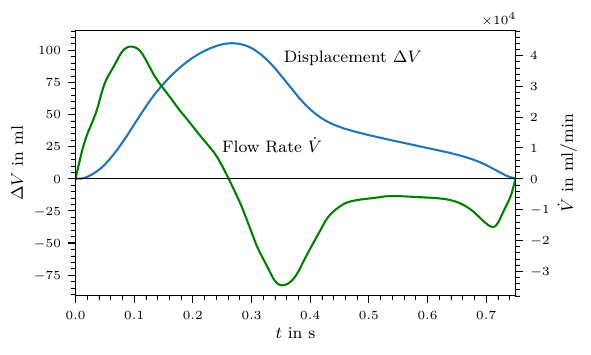}
    \vspace{-6mm}
    \caption{\label{fig:BC} Cardiac piston pump displacement $\Delta V$ (blue) and volume flow rate $\dot V$ (green) for one Pinhole-L cycle.}
\end{figure}

In the experiment, the transient flow was driven by a ViVitro Labs cardiac piston pump operating at 80 beats per minute, representing a $0.75\,\mathrm{s}$ cycle period. In the computational model, the pump was defined as a transient mass flow inlet across a fixed circular area (magenta patch in Fig.~\ref{fig:mesh},b). For each MROP in the physical experiment, the pump displacement waveform was scaled to generate a uniform physiological peak systolic pressure of $120\,\mathrm{mmHg}$ in the LV. Larger MROP orifices presented lower flow resistance and required a higher prescribed mass flow to reach the target pressure. The scaled volume displacement was differentiated over time, converted to mass flow, and prescribed via a time-interpolated table. The displacement waveform and resulting flow rate are shown in Fig.~\ref{fig:BC} for the Pinhole-L case ($\Delta V=106.6\,\mathrm{ml}$). 

The free surface in the upper reservoir is simulated as an outlet at a constant pressure ($0\,\mathrm{Pa}$) 
(orange patch in Fig.~\ref{fig:mesh},b), while all internal surfaces were set as no-slip boundaries. Boundary conditions for the turbulent kinetic energy ($k$) and specific dissipation rate ($\omega$) at both the inlet and outlet were specified using a turbulence intensity and turbulent viscosity ratio of zero. The internal fluid domain was similarly initialized with zero velocity, pressure, and turbulence fields. Due to this cold start, the simulations were run for at least two full cardiac cycles to establish a developed periodic flow regime, with only the latest cycle being used for data extraction. 
While minor cycle-to-cycle fluctuations and slow convergence of fine-scale flow details naturally persist, the macroscopic flow characteristics, peak velocity magnitudes, and global jet topology are fully established and stable by the second cycle.

\subsubsection{\label{sec:extraction} Training Data Extraction}

Internal fields were exported every $2.5\,\mathrm{ms}$ of simulation time as EnSight Gold Case files, generating 300 discrete snapshots, from the latest CFD cycle of each MROP. The included fields were chosen as the three velocity components ($u,v,w$), pressure ($p$), turbulent viscosity ($\nu_t$), specific dissipation rate ($\omega$) and turbulent kinetic energy ($k$). This data is stored on the vertices of the polyhedral mesh cells, resulting in more data points than the total cell count. This point cloud indirectly retains the fluid domain geometry including the domain boundaries with walls inherently identified by zero velocities through no-slip boundary condition. Point density also varies due to the localized mesh refinements.
Alongside these fields, virtual pressure probes were exported at the same coordinates as the physical sensor locations (see Sec.~\ref{sec:experiment}). Defining the origin of Cartesian coordinates at the center of the orifice with the $x$-axis oriented along the regurgitant jet direction (Fig.~\ref{fig:slotlbent},a), these points are $\boldsymbol{x}_\mathrm{LV}=(-11.825, -3.680,\ 0.0)\,\mathrm{cm}$ and $\boldsymbol{x}_\mathrm{LA}=(0.986,\ 0.008, -4.0)\,\mathrm{cm}$.

The exported data was subsequently cropped spatially to minimize network training overhead. The 3D domain was truncated to omit points where $y>6\,\mathrm{cm}$ or $z<-5\,\mathrm{cm}$. This limited the field to the relevant portions of the LV and LA, omitting the aorta, reservoir, and long connection to the pump.

\subsection{Experimental Dataset}
\label{sec:experiment}

\begin{figure*}
    \centering
    \includegraphics[]{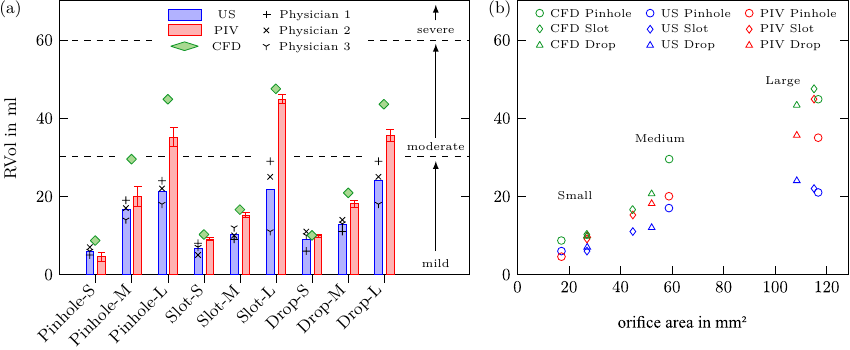}
    \caption{\label{fig:rvol} Regurgitation volume (RVol) of MROPs measured via ultrasound (US) by three physicians and PIV with uncertainty confidence intervals. CFD results (in green) are added to the data from Leister \textit{et al.}~\cite{Leister2025}}
\end{figure*}

To qualify the synthetic CFD training database and evaluate the performance of the proposed framework under real-world conditions, we utilized an existing experimental dataset acquired from the physical simulator.\cite{Leister2025} The measurements were captured using two-component two-dimensional particle image velocimetry (2D2C PIV). For these, the working fluid was seeded with $20\,\mathrm{\mu m}$ polyamide particles. The atrial flow field was illuminated along specified $x$-$y$-planes using a double-pulsed Nd:YAG laser ($532\,\mathrm{nm}$) and recorded via an sCMOS camera through a circular window. The window shape is visible in Fig.~\ref{fig:mesh},b around the ``LA'' designation.

The subset of data used from the existing work by Leister~\textit{et~al.}~\cite{Leister2025} are the recordings which resolve the phase-averaged transient flow. 
Images were captured at up to 50 discrete phase positions across the cardiac cycle, with 100 double-images averaged per phase, which increases statistical significance and aims to filter out chaotic turbulent fluctuations and measurement noise. 
Additionally, this data effectively represents a cyclic-steady state that is conceptually similar to the carried-out URANS-based simulations.
These cardiac-phase resolved measurements covered four distinct MROP configurations: Pinhole-L, Slot-L, Drop-XL, and EccJet. Apart from the EccJet case, only the center plane at $z=0\,\mathrm{mm}$ was recorded, resulting largely in an absence of 3D time-resolved data that could be used as ground truth. The EccJet geometry is also the only configuration featuring a rigid rather than flexible, dynamically deforming orifice plate.

In the previous work, in addition to planar velocity fields, the overall regurgitation volume (RVol) was quantified using transesophageal echocardiography (TEE). Three experienced physicians independently performed these evaluations using different clinical ultrasound systems (Epiq Cvxi, Epiq 7c, and iE33) via the conventional flow convergence (PISA) method.\cite{Leister2025} 
Using the available data, figure~\ref{fig:rvol} compares the resulting regurgitation volumes across the three modalities: ultrasound, PIV, and the synthetic URANS simulations. The comparative results show good agreement between CFD and experiments, with the CFD model capturing the clinical severity trends across different MROP geometries and sizes, while consistently overestimating the regurgitation volume. A more exhaustive spatiotemporal comparison including phase-resolved PIV velocity fields, mass flow, and pressures measurements is provided in Appendix~\ref{sec:validation}. Overall, despite the quantifiable deviations, the CFD database proved reasonably accurate and successfully captured the key topological and transient flow features required for training.


\subsection{\label{sec:methodology:ssec:architecture}DeepONet Architecture}

\begin{figure*}[]
    \centering
    \includegraphics[width=0.8\textwidth]{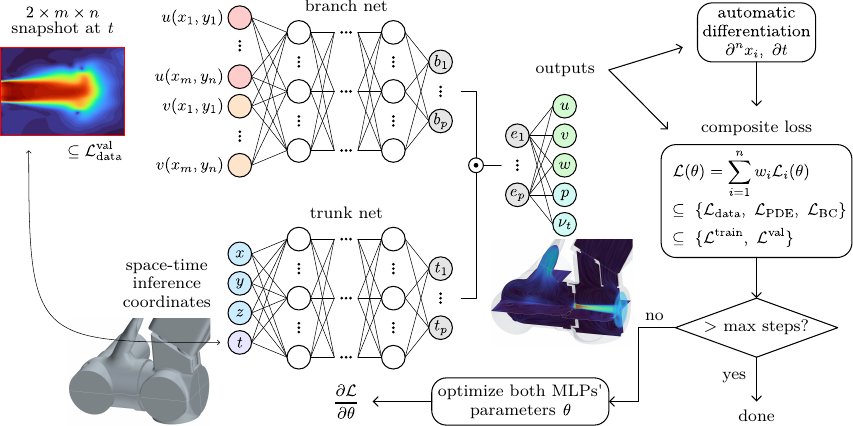} 
    \caption{\label{fig:deeponet_schematic} Schematic of the DeepONet architecture for 3D transient flow reconstruction. The branch net receives the constrained 2D2C velocity snapshot in a flattened format. The same snapshots are additionally trained on for Test-Time Adaptation (TTA). The current snapshot's time-step is the trunk net input $t$, alongside the spatial inference coordinates. Inputs and outputs are non-dimensional (notation $\square^*$ omitted for legibility).}
\end{figure*}

To reconstruct the 4D flow fields, a Deep Operator Network (DeepONet) approach is employed. \cite{37} Unlike standard neural networks that map coordinate points to solutions for a single case, DeepONets learn a general parametric mapping between input and output functions. The architecture consists of two parallel subnetworks: a branch network and a trunk network, as shown in Fig.~\ref{fig:deeponet_schematic}. 

The branch network is responsible for ingesting the conditional input function. In this work, this input is defined as the flattened, spatially masked 2D velocity slice $(u, v)$ extracted from the $x$-$y$-plane at $z=0\,\mathrm{cm}$ and a specific time-step $t$, directly mimicking the restricted field of view (FOV) obtainable via 2D2C PIV measurements. This constitutes a snapshot-based approach, where inferring the full 3D volumetric field at any specific instant $t$ is tied to the availability of a corresponding 2D branch input from that same time-step.  

To format this conditional input, the extracted slice is interpolated onto a $119 \times 102$ equidistant grid spanning the coordinates $x \in [0.05, 6.9]\,\mathrm{cm}$ and $y \in [-2.2, 2.2]\,\mathrm{cm}$. For the Slot-L-Bent case, this extraction area was shifted in the $x$-direction by $6.4\,\mathrm{mm}$, since it would otherwise partly capture flow inside the LV (Fig.~\ref{fig:slotlbent}).
Because the physical PIV camera setups varied across experiments and utilized a circular optical window, this rectangular grid is subjected to a polygonal mask to ensure a common FOV across all datasets: Linear cutoffs are applied at $-30^\circ$ and $60^\circ$ in the distal corners, yielding a standardized, flattened input vector comprising exactly $N = 10,738$ discrete sampling locations for every case, and thereby $2N$ branch input neurons to accommodate both $u$ and $v$. 
While the specific selection of these observation locations is ultimately arbitrary, maintaining a consistent ordering is necessary: since the branch network is supplied only with the velocities and receives no information regarding the coordinates for these values, it relies on the fixed mapping to learn the relationship between the 2D slice and the resulting volumetric flow. 

Concurrently, the trunk network ingests the continuous spatiotemporal inference coordinates $(x, y, z, t)$ where the solution is desired. Both subnetworks map their respective inputs to a latent representation of dimension $p$. The outputs of both networks are then merged via element-wise multiplication in this latent space. Finally, a linear projection layer maps this merged latent vector to the targeted output quantities.

\hyphenation{Phy-sics-Ne-Mo}
The model is implemented using the NVIDIA PhysicsNeMo Symbolic framework (v1.5.0, formerly NVIDIA Modulus Symbolic).~\cite{44, 45} Both subnetworks are constructed as Multilayer Perceptrons (MLPs). Based on hyperparameter evaluations, a high-capacity architecture consisting of 9 hidden layers with 640 neurons each was selected for both. Adaptive SiLU (Swish) activation functions are utilized across all hidden layers to capture the highly non-linear nature of the regurgitant jets.~\cite{11, 25} In this work, the latent dimension $p$ is equal to the hidden layer width, resulting in $p=640$. 

The complete dimensional mapping through the network can therefore be summarized as follows: the branch input ($\mathbb{R}^{21,476}$) and trunk input ($\mathbb{R}^4$) are independently mapped to latent space ($\mathbb{R}^{640}$), combined via element-wise multiplication ($\mathbb{R}^{640} \odot \mathbb{R}^{640} \rightarrow \mathbb{R}^{640}$), and finally reduced via linear projection to the five target output flow fields ($\mathbb{R}^5$). In total, this configuration yields 21,147,522 trainable artificial neural network parameters, collectively denoted as $\theta$.

\subsection{\label{sec:methodology:ssec:strategy}Training Strategy}
To improve training convergence and consistency by bounding parameter magnitudes near unity, all physical quantities $\alpha$ are non-dimensionalized as $\alpha^* = \alpha/\alpha_c$ before being supplied to the network. The characteristic length $l_c=0.1\,\mathrm{m}$ roughly normalizes the spatial domain, $t_c=0.75\,\mathrm{s}$ reflects one cardiac cycle period, and the remaining scales correspond to the global maximum values observed across the entire CFD dataset: $u_c=4.988\,\mathrm{m/s}$, $p_c=13439.05\,\mathrm{Pa}$, and $\nu_{t,c}=3.92\cdot10^{-4}\,\mathrm{m^2/s}$.

The training of the DeepONet framework follows a two-step ``Test-Time Adaptation'' (TTA) paradigm to maximize accuracy for unseen configurations. The optimization is driven by minimizing a composite loss function $\mathcal{L}(\theta)$ using the AdamW optimizer with exponential decay of the learning rate where the initial rate of 0.001 is multiplied by 0.95 every 4000 steps. \cite{36}

In the first step (pre-training), the network is optimized on the full prior-knowledge CFD dataset to learn the generalized operator. This phase relies purely on a standard data-driven loss ($\mathcal{L}_{\mathrm{data}}^{\mathrm{train}}$), comparing the network's predictions to the full 3D transient CFD fields across 50,000 training steps, which required approximately 12 hours of computation time (RTX 6000 Ada GPU). Because the network is trained on the URANS dataset, it by extension attempts to predict the phase-averaged flow fields (e.g., $p^*=\overline{p^*}(x,t)$) rather than instantaneous, chaotic turbulent fluctuations. 

In the second step (hybrid fine-tuning), the pre-trained weights are adapted to the specific, unseen validation case. This step acts as a rapid calibration, utilizing phase-averaged pressure measurements, one each in the left atrium ($\boldsymbol{x}_\mathrm{LA}$) and left ventricle ($\boldsymbol{x}_\mathrm{LV}$), and, more importantly, the sparse 2D PIV velocity slice of the target case ($\mathcal{L}_{\mathrm{data}}^{\mathrm{val}}$) in addition to the ongoing training distribution ($\mathcal{L}_{\mathrm{data}}^{\mathrm{train}}$). This fine-tuning process is highly efficient, requiring only 400 additional iterations, taking from 10 minutes to 1.5 hours depending on the case. Iteration counts for both steps were determined manually based on initial trial runs to balance optimal convergence with the prevention of overfitting. An automated stopping criterion was not implemented in this study due to the complexity of balancing competing performance indicators. Instead, termination was guided by monitoring quantitative errors against available ground truth data for both the training and validation cases, alongside qualitative visual assessments of the predicted flow topologies. The reported computation times reflect a development environment with frequent, costly validation steps and do not represent a fully optimized framework.

The mathematical formulation of the data loss utilizes the Sum of Squared Errors (SSE, Appendix~\ref{app:SSE}) across all output variables: the velocity components ($u, v, w$), static pressure ($p$), and the kinematic eddy viscosity ($\nu_t$). To effectively balance the gradients of these competing loss terms, Neural Tangent Kernel (NTK) based dynamic weighting is employed, with updates to the weights being calculated every 100 training steps. \cite{58} 

Notably, the turbulent viscosity $\nu_t$ is included as a direct network output. This allows the framework to close the Reynolds-averaged Navier-Stokes equations without the complexity of calculating full transport PDEs for $k$ and $\omega$. This quantity, like the velocity components and pressure, can be supervised directly in the case of (U)RANS simulation data availability. \cite{47} While the inclusion of physical constraints via PDE residual losses ($\mathcal{L}_{\mathrm{PDE}}$) was extensively tested during development, amounting to Physics-Informed DeepONets, they were ultimately omitted from the final configuration. The purely data-driven fine-tuning proved more robust; including PDE loss terms in the highly complex 3D transient domain induced over-smoothing and significantly increased computational costs without yielding tangible accuracy gains.

%% file: sections/results.tex
\subsection{Validation on Synthetic Data}

To assess the framework's predictive capabilities and its ability to handle out-of-distribution inputs, the DeepONet was first evaluated on the unseen synthetic CFD data. 
Firstly, the larger orifice Drop-XL is discussed. Since the training database only included orifice sizes up to a `Large' designation, predicting the `Extra-Large' geometry represents a challenging extrapolation task. 

\begin{figure*}[]
    \centering
    \includegraphics[width=\textwidth]{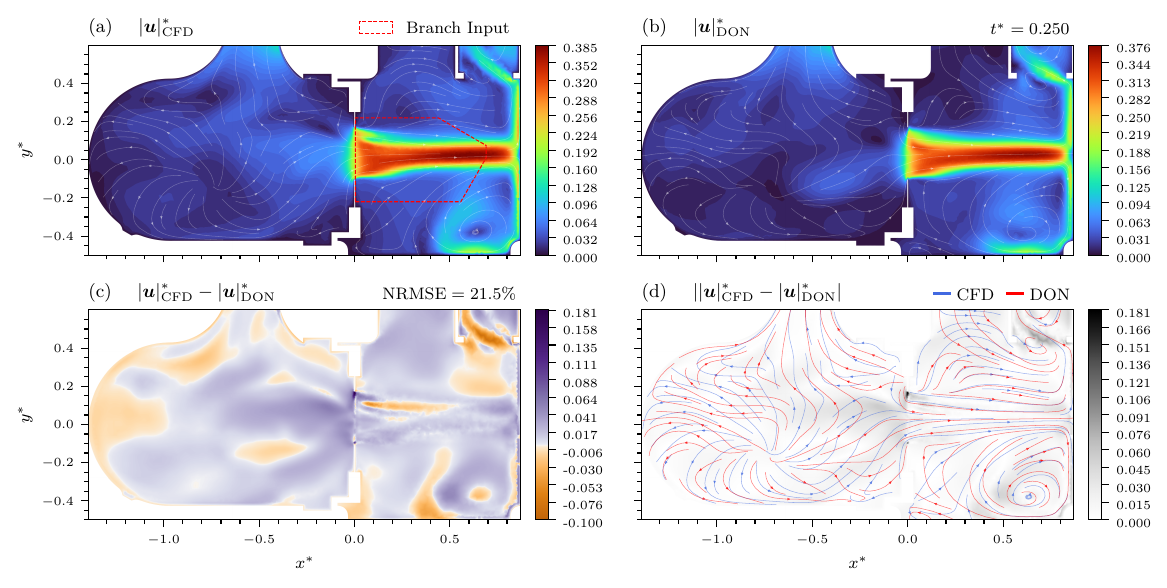}
    \vspace{-8mm}
    \caption{Velocity magnitude distribution at $z^*=0$ and $t^*=0.25$ for the Drop-XL orifice. (a) Reference CFD solution, (b) DeepONet prediction. (c) Signed, non-linearly displayed and (d) absolute error magnitude with streamline comparison.}
    \label{fig:val}
\end{figure*}

Fig.~\ref{fig:val} visualizes the spatial reconstruction capabilities of the hybrid model during the systolic phase at the time instant $t^*=0.25$. What is referred to as the `DeepONet' prediction, such as in these velocity magnitude comparisons, includes the MROP-specific TTA training-step, unless specified otherwise. The framework reconstructs the regurgitant jet with high fidelity, accurately matching both the magnitude and the spatial extent of the CFD reference.

To precisely evaluate these reconstructions, the signed, non-linearly scaled error (Fig.~\ref{fig:val},c) highlights even subtle deviations across the domain. The DeepONet slightly underpredicts the velocity magnitude (indicated by dark purple regions) in the flow approaching the orifice from the LV, as well as in the left upward path connecting the LA to the reservoir. 
Minor temporal and magnitude deviations are also visible in the vortex that rolls down the LA wall; the network predicts this structure to be slightly more advanced along the wall and less intense than the CFD reference. In contrast, the velocity is too high in the right upper path to the reservoir.  Within the regurgitant jet itself, the network occasionally smooths over fine localized features. For example, near the upper edge of the orifice plate, the CFD solution features a subtle dip in velocity magnitude that the TTA does not entirely resolve, resulting in a localized band of overprediction (orange). However, because this specific area falls within this additionally supervised area, the ground truth data could theoretically be substituted post-inference, making this localized in-plane error less critical.

Plotting the absolute error linearly alongside the superimposed streamlines (Fig.~\ref{fig:val},d) highlights the regions of highest deviation. The most prominent absolute errors occur directly at the boundaries of the orifice ($x^*=0$). Because the baseline model struggles to extrapolate to the Extra-Large orifice size, it initially predicts a narrower jet. The TTA supervision window begins just downstream of the orifice and does not reach into the opening itself. This reveals an important limitation of the adaptation process: the spatially localized planar correction fails to propagate significantly upstream. Consequently, an uncorrected gap remains between the physical orifice wall and the beginning of the TTA zone, resulting in concentrated regions of underpredicted velocity. This effect is especially visible at the upper boundary because, in this $z^*=0$ cross-section, the narrowly tapering upper tip of the drop shape extends further upward than the rounded bottom, exposing a larger area of uncorrected flow.

Despite these localized absolute errors, the streamline comparison (Fig.~\ref{fig:val},d) demonstrates strong overall topological agreement. Crucially, the prediction respects the complex geometric boundaries in the high-momentum regions; the jet interacts naturally with the atrial walls, rolling off the internal surfaces to the sides, which results in low error magnitudes along the fluid-solid interfaces. While these primary jet dynamics are captured excellently, exact flow topology matching across the entirety of the complex 3D domain proves challenging. Misalignments in the velocity vectors are largely confined to low-velocity regions, such as the lower-left corner of the LA, the aforementioned upward path to the reservoir, and outside the core of the circular structure in the LV. In these quiescent zones, even minor absolute velocity deviations can significantly alter the resulting streamline trajectories, leading to visible topological differences despite low scalar error magnitudes.

The impact of the Test-Time Adaptation (TTA) step becomes further evident when analyzing the temporal development of the jet. Physically, the precise moment of jet initiation varies depending on the specific MROP geometry. As demonstrated by the flow rate comparison in Appendix~\ref{sec:appresults} (Fig.~\ref{fig:slotslm}), the bulk transvalvular flow reversal does not align uniformly with the driving pump waveform; this was also observed in the PIV experiment. Although the scaled pump boundary conditions share identical temporal zero-crossings, larger orifices require higher total mass flows to reach the target $120\,\mathrm{mmHg}$ ventricular pressure. This imparts greater inertia to the bulk fluid, delaying its deceleration and subsequent flow reversal in the cardiac cycle. Conversely, the lower bulk inertia associated with smaller, higher-resistance orifices allows the regurgitant jet to initiate earlier.

From a theoretical standpoint, the DeepONet might ideally learn to capture this timing without secondary calibration. For that, because the 2D measurement slice is fed into the branch network, the model would have to map those values directly to the corresponding spatiotemporal inference coordinates. Essentially, the known planar data would pass through the network at that specific location, embedding the sparse observational input as part of the volumetric output solution.

In practice, however, the present baseline model does not achieve such an exact mapping and is potentially negatively affected by the variety in jet initiation timing during pre-training, frequently positioning the jet pulse temporally ahead of the ground truth. As illustrated for the Drop-XL case in Appendix~\ref{sec:appresults} (Fig.~\ref{fig:preTTA}), the baseline prediction prematurely extends the jet to the atrial wall while paradoxically exhibiting a lower peak velocity magnitude than the CFD reference, indicating a physically inconsistent development. The hybrid TTA step successfully addresses this by accurately retracting the jet front within the supervised $z^*=0$ plane to match the sparse observations, while also successfully recovering the correct velocity magnitude across the core of the jet (Appendix Fig.~\ref{fig:postTTA}). 
However, while the magnitude within the jet core improves, the predicted jet remains distinctly narrower than the CFD ground truth. In both figures' spatial error maps (c, d), this discrepancy manifests as prominent ``streaks'' originating at the orifice plate wall. They are narrow, elongated bands of high error flanking the core jet region, caused by the width of the jet being underpredicted. This specific error stems from the extrapolation to an orifice size larger than encountered during training. 
Similarly, the spatial retraction of the jet front does not propagate far perpendicularly into the unconstrained $z$-direction of the domain. Consequently, while the center plane is adjusted closer to the near-zero velocity between the jet front and the wall, the outer fluid layers remain in their original, temporally advanced position. This leaves erroneously elevated velocities in the outer $z$-layers, creating a disjointed flow appearance in the cross-section.

During the diastolic phase, the DeepONet continues to successfully capture the overall topology, with streamlines largely matching the ground truth. This diastolic jet is a non-physiological remnant of the fixed MROP experimental design, but it provides additional insight for assessing the model’s robustness. 
Because the planar TTA supervision fails to more meaningfully propagate into the depth of the domain or laterally, the previously defined underprediction streaks persist throughout this phase as well. Driven by the same geometric extrapolation, these narrow error bands span across both the unconstrained off-center planes in the LA ($x^*>0$, $z^*\neq0$) and the entirety of the LV ($x^*<0$), where no direct supervision is available.

\begin{figure*}[] 
    \centering
    \includegraphics[width=\textwidth]{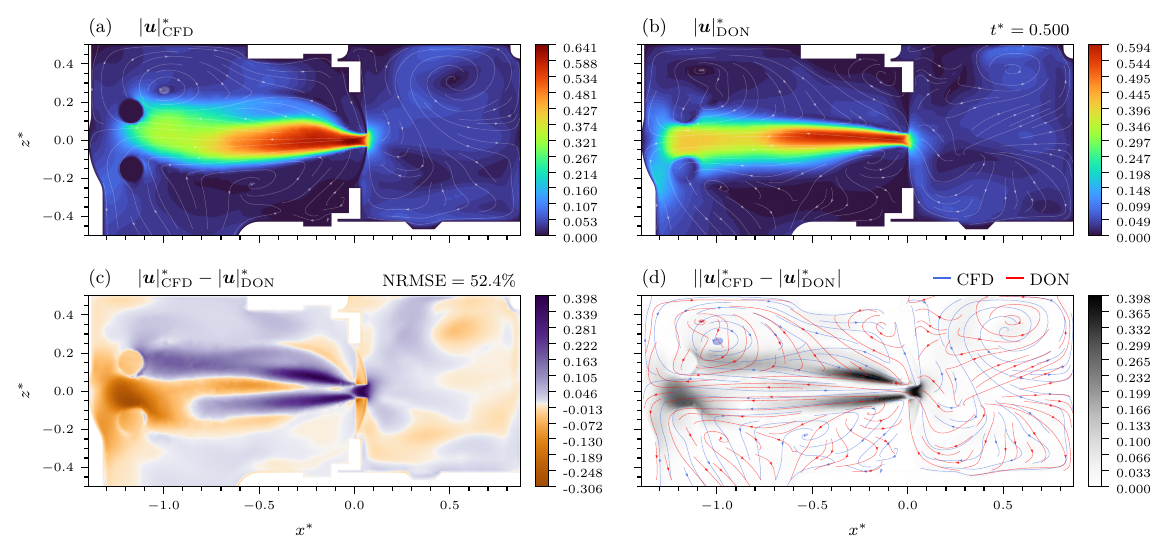}
    \vspace{-8mm}
    \caption{Velocity magnitude distribution at $y^*=0$ and $t^*=0.5$ for the Slot-L-Bent orifice. (a) Reference CFD solution, (b) DeepONet prediction. (c) Signed, non-linearly displayed and (d) absolute error magnitude with streamline comparison.}
    \label{fig:val_y}
\end{figure*}

Similar error streaks can be observed in the second validation case: the Slot-L-Bent geometry. To account for the physical displacement of the orifice opening caused by the spherical wall, the branch network input was extracted from a location shifted in the positive $x$-direction (see Fig.~\ref{fig:slotlbent} and Section \ref{sec:methodology:ssec:architecture}). However, the sparse supervision was kept anchored to the original coordinate system. This positioning strategy ensures that the DeepONet infers the jet initiation at a location consistent with its prior training distribution. 
During the systolic phase, this strategy yields an overall regurgitant jet reconstruction that matches the ground truth well. Both the accuracy successes and the specific local limitations resemble those observed in the Drop-XL case. The hybrid model successfully captures the primary jet magnitude and boundary-following streamlines, while some deviations remain in low-velocity regions such as the upper reservoir path, alongside a slightly temporally advanced vortex roll-off in the LA. One minor contrasting detail occurs in the flow approaching the orifice from the LV, which exhibits a slight overprediction in velocity magnitude rather than an underprediction. However, while the bulk jet is well reconstructed, the inherent geometric mismatch between the network's unshifted coordinate anchor and the physically bent orifice creates a persistent, localized error spike in the gap between the usually flat orifice plate and the actual jet origin of the bent plate.

Furthermore, the spatial constraints of the adaptation process again become evident in the jet width. As shown in Fig.~\ref{fig:val_y} for the diastolic phase ($t^*=0.5$), distinct narrow bands of high error (streaks) are visible along the jet boundaries throughout the domain, again indicating an underpredicted, narrower jet shape. Streamline comparisons during this specific phase further show a breakdown in topological agreement. In the LA, the overall flow field deviates significantly from the ground truth, where the prominent vortex located in the top right of the CFD reference is barely resolved by the network. In the LV, the CFD reference presents a challenging asymmetric topology, featuring a single vortex near the jet tip, that the DeepONet fails to predict with a well-defined center. Furthermore, the ground truth flow in the surrounding low-velocity regions of the LV exhibits fine topological details. The network struggles to match these low-momentum features, resulting in largely misaligned predicted streamlines in these unconstrained areas. Because the standard, flat Slot-L geometry is included in the training distribution, these errors are a result of the network struggling to fully resolve the novel bent orifice plate accompanied by the input coordinate shift and altered jet shape. In this out-of-distribution scenario, the hybrid TTA step yielded only minimal quantitative improvement over the baseline prediction, highlighting the framework's reliance on the initial training distribution's coverage.

\begin{figure*}
    \centering
    \includegraphics[width=\textwidth]{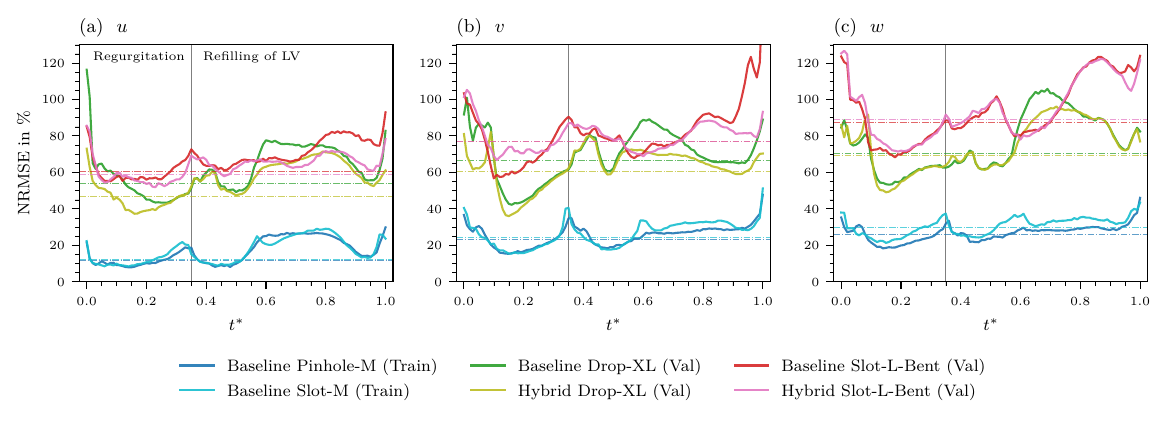}
    \vspace{-8mm}
    \caption{Temporal evolution of NRMSE for the three velocity components over the cardiac cycle. The vertical line marks the moment of flow reversal, indicating the onset of the diastolic jet forming in the LV. Includes pre-trained baseline and fine-tuned hybrid results for validation cases and select training cases. Horizontal dashed lines indicate the respective aggregate values computed over the entire period.}
    \label{fig:overtime}
\end{figure*}

To quantify the framework's performance across the entire cardiac cycle, Fig.~\ref{fig:overtime} illustrates the temporal evolution of the Normalized Root Mean Square Error (NRMSE, Appendix~\ref{app:NRMSE}) for the three velocity components ($u, v, w$). As expected, the network achieves largely consistent and accurate predictions on the training geometries (e.g., Pinhole-M), with instantaneous NRMSE values generally ranging between 10\% and 40\% throughout the cycle. 

In contrast, the validation cases exhibit higher error magnitudes, between roughly 40\% and 100\%. For the Drop-XL case, the horizontal lines illustrate a modest overall reduction in aggregate NRMSE following the hybrid adaptation step (see Appendix~\ref{app:NRMSE} for the calculation and its difference from a simple mean).

However, the temporal traces reveal that this improvement varies across the cardiac cycle. Some of the error reductions for both MROPs occur during systole as seen when comparing the baseline to the respective hybrid results in Fig.~\ref{fig:overtime} (left of the vertical line), but comparatively substantial reductions are also observed during the diastolic refilling phase (right of the vertical line), where velocity magnitudes within the supervised 2D2C window are considerably lower. This indicates that the TTA step does not strictly rely on high velocity magnitudes to guide the network.

Interestingly, this adaptation also yields intermittent error reductions in the $w$-component for the Drop-XL case (Fig.~\ref{fig:overtime},c), despite this velocity component receiving no additional supervision in the hybrid step. However, it remains inconclusive whether this reflects a genuine, holistic correction of the 3D flow state, or merely unconstrained, localized network adjustments that coincidentally yield a lower aggregate scalar metric for these select phases. Regardless of the reason for this error reduction, the previously described disjointed volumetric structures observed in the $z$-direction obviously persist to an unsatisfactory degree.

Another notable feature across the error traces is a localized peak corresponding to the moment of flow reversal, marked by the vertical line. During this highly transient phase, the dissolution of the systolic jet and the initiation of the diastolic jet create a multidirectional flow topology. Because the network might struggle to perfectly synchronize this temporal transition, phase misalignments can result in elevated squared error where predicted and true velocity vectors briefly point in opposite direction. While this reversal peak is distinct for the well-predicted training cases, it is somewhat less pronounced in the baseline and hybrid validation models, as their error traces are already elevated throughout the cycle due to the overarching challenges of geometric extrapolation.

Despite the localized improvements observed in both validation cases, there are also periods where the baseline and hybrid error traces largely overlap. Furthermore, examining the aggregate scalar metrics (horizontal dashed lines) reveals inconsistent benefits across the models. As previously mentioned, the Drop-XL case achieves a modest overall reduction in aggregate NRMSE following adaptation, but the Slot-L-Bent case actually exhibits a higher aggregate error for the $v$ and $w$ components. The temporal evolution for Slot-L-Bent demonstrates that the hybrid prediction yields a slightly higher NRMSE than the purely data-driven baseline during several intervals, notably including the systolic phase. This indicates that while the fine-tuning step successfully corrects specific topological features, such as the shifted jet origin resulting from the bent orifice plate, the improvement may not be reflected in aggregate scalar error metrics.

\begin{figure}[]
    \centering
    \includegraphics[width=1\linewidth]{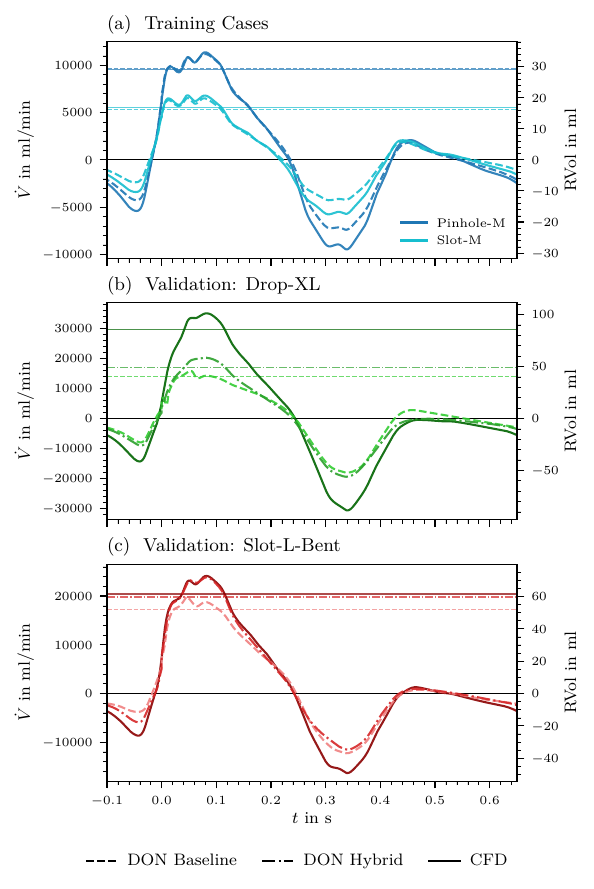}
    \vspace{-7mm}
    \caption{\label{fig:mass_flow_comparison_paper} Mitral volume flow and integrated regurgitation volume (RVol, horizontal lines) for select MROPs. Comparison of the CFD reference (solid) with baseline (dashed) and fine-tuned hybrid (dash-dotted) DeepONet predictions: (a) Pinhole-M and Slot-M training cases (baseline only), (b) Drop-XL and (c) Slot-L-Bent validation cases.}
\end{figure}

These findings highlight a characteristic limitation of the current adaptation process: the sparse planar supervision does not consistently cause accuracy improvements throughout the entire fluid domain and cycle, but instead frequently remains localized. To assess how these localized topological deviations translate into a clinically established metric, the volume flow rate and integrated regurgitation volume (RVol) were calculated from the network predictions. The flow was extracted through a plane positioned $0.5\,\mathrm{mm}$ downstream of the orifice, sized and shaped to slightly overshoot the specific MROP opening. RVol was defined by integrating the positive flux during the systolic phase. Fig.~\ref{fig:mass_flow_comparison_paper} compares the resulting temporal traces and scalar volumes.

For the training geometries, represented by Pinhole-M and Slot-M (Fig.~\ref{fig:mass_flow_comparison_paper},a), the purely data-driven baseline prediction matches the CFD reference closely during systole. A noticeable difference occurs during the diastolic flow reversal, where the network predicts a lower magnitude trace. Firstly, because the neural network output does not strictly satisfy mass conservation, the computed volumetric flux can be sensitive to the exact position of the evaluation plane. Secondly, during the diastolic refilling phase, the flow behaves as a diffuse sink approaching the orifice. It is possible that fluid partially bypasses the boundaries of the integration plane due to the position being $0.5\,\mathrm{mm}$ into the LA, leading to the observed deviation.

Applying this calculation to the validation cases illustrates differing challenges of out-of-distribution inference. For the oversized Drop-XL geometry (Fig.~\ref{fig:mass_flow_comparison_paper},b), the baseline model underestimates the mass flow and resulting RVol significantly. This deficit corresponds to the underpredicted jet width discussed previously. Because volume flow scales with the cross-sectional area of the jet core, the network's inability to extrapolate to a significantly larger orifice area limits the integrated volume. The localized TTA step provides only a modest correction.

In contrast, the Slot-L-Bent configuration (Fig.~\ref{fig:mass_flow_comparison_paper},c) demonstrates a stronger recovery. While the baseline prediction initially underestimates the mass flow, yielding an RVol of $51.85\,\mathrm{ml}$, the hybrid adaptation step results in a closer match of $59.58\,\mathrm{ml}$ compared against the CFD reference of $61.48\,\mathrm{ml}$. Although this geometry shares the same underlying orifice area as the standard Slot-L training case, the bent orifice plate generates a broader jet topology than the tapering jet of the flat plate (Fig~\ref{fig:planerender_slotlphasebent}). The result here suggests that the TTA step can be effective at correcting the flow trajectory and compensating for the narrower baseline prediction for RVol, bearing in mind that having the identically driven Slot-L baseline in the training distribution likely contributed to this success.

Ultimately, this synthetic validation establishes the specific capabilities and limits of the single-plane TTA strategy. The TTA training step proves capable of correcting in-plane features, such as retracting a temporally premature jet front or recovering the integrated volume flow for geometric variations of known defect sizes. However, these planar corrections consistently fail to propagate into the unobserved volume. When the baseline prediction requires substantial corrections, the localized supervision can result in disjointed 3D structures and an uncorrected deficit in bulk flow. These findings demonstrate that the TTA fine-tuning step, as implemented in this work, is insufficient on its own to guarantee satisfactory results across the unconstrained 3D domain.

\subsection{Application to Experimental Data}

Moving beyond validation on synthetic data, the framework was applied to real-world in-vitro measurements. This introduces several constraints, primarily the absence of a full 3D volumetric ground truth, inherent measurement noise, and a domain shift caused by the flexible orifice foil used in the experiment versus the rigid walls modeled in the CFD training distribution. Furthermore, the available experimental PIV data contains fewer time-steps, providing only 25 to 50 snapshots per case compared to the 300 in the synthetic dataset. Due to the snapshot-based nature of the framework, this restricts the 3D volumetric reconstructions solely to those specific recorded instants. The temporal sparsity of frames may also affect the overall reconstruction quality, as less case-related information is available to fine-tune on.

\begin{figure}[]
    \centering
    \includegraphics[width=1\linewidth]{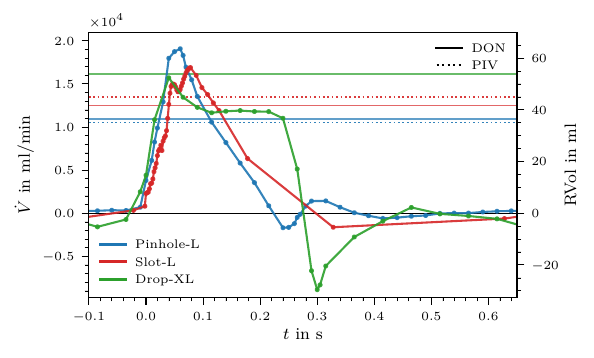}
    \vspace{-7mm}
\caption{\label{fig:mass_flow_comparison_exp_paper} DeepONet predictions for mitral volume flow and integrated regurgitation volume (RVol, horizontal lines) driven by experimental PIV input. PIV-derived reference RVol (dotted) for Pinhole-L and Slot-L available from Fig.~\ref{fig:rvol} and corresponding paper.\cite{Leister2025} Dots mark the discrete phase positions captured by the phase-averaged PIV.}
\end{figure}

The framework's macroscopic predictive capability was first evaluated by computing the transient volume flow rate and the resulting integrated regurgitation volume (RVol). Using the same evaluation methodology applied to the synthetic data, Fig.~\ref{fig:mass_flow_comparison_exp_paper} presents the predictions driven by the sparse sequence of experimental 2D2C PIV snapshots. For the Pinhole-L and Slot-L configurations, the network predicts RVol values of $36.41\,\mathrm{ml}$ and $41.67\,\mathrm{ml}$. These align closely with the physical PIV-derived references of $35.1\,\mathrm{ml}$ and $44.9\,\mathrm{ml}$, respectively. No PIV-derived RVol is available for the Drop-XL case. 

The predicted temporal traces also exhibit distinct transient behaviors. The Pinhole-L mass flow features a relatively sharp descent during late systole, while the Drop-XL prediction displays a pronounced, extended plateau. This plateau reflects the hardware-induced clipping of the physical cardiac pump waveform documented during the experiment (detailed in Appendix~\ref{sec:validation}). The network successfully infers this atypical driving dynamic relying solely on the sparse sequence of 2D velocity fields. 

\begin{figure}
  \centering

    \includegraphics[width=\linewidth]{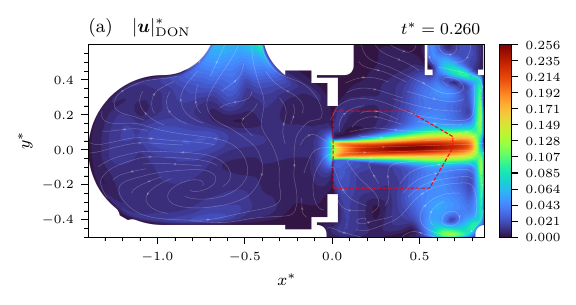}
    \vspace{-6.5mm}
      
    \includegraphics[width=\linewidth]{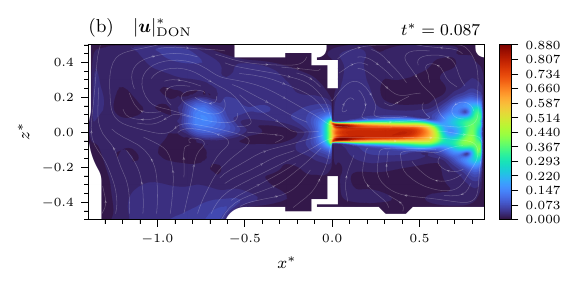}
    \vspace{-6.5mm}
    
    \includegraphics[width=\linewidth]{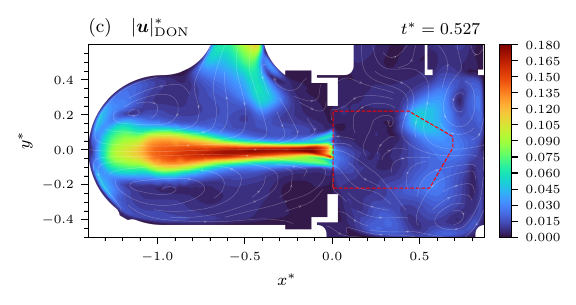}
    \vspace{-8mm}

  \caption{DeepONet velocity magnitude predictions for the Pinhole-L orifice based on experimental PIV input. (a) $x$-$y$-plane at $z^* = 0$ and $t^* = 0.26$, (b) $x$-$z$-plane at $y^*=0$ and $t^* = 0.087$, (c) $x$-$y$-plane at $z^* = 0$ and $t^* = 0.527$.}
  \label{fig:pinholelphase}
\end{figure}

However, as demonstrated during the synthetic validation, accurate integrated bulk flow can obscure localized topological failures. The spatial and structural coherence of these experimental predictions must therefore be examined in detail. The spatial analysis begins with the Pinhole-L configuration, which exhibited the strongest agreement during validation of the CFD simulations (see Appendix~\ref{sec:validation}). Fig.~\ref{fig:pinholelphase},a demonstrates that the fine-tuned DeepONet predicts a physically plausible flow topology during the systolic phase ($t^*=0.26$). The regurgitant jet extends naturally to the atrial wall and correctly rolls off the geometry. However, limitations regarding volumetric coherence persist. Fig.~\ref{fig:pinholelphase},b illustrates the $x$-$z$-plane at an earlier time-step ($t^*=0.087$), revealing a premature disjointed connection of the jet front to the wall. The TTA correction within the $z^*=0$ plane fails to propagate laterally, leaving the outer 3D jet layers uncorrected in their advanced temporal position. Since the supervised area does not reach all the way to the wall, even the center plane is inaccurate here. 
Furthermore, the diastolic phase (Fig.~\ref{fig:pinholelphase},c) exhibits non-physical artifacts. The prediction displays a strong central jet core surrounded by a large, diffuse region of weaker velocity magnitude, which implausibly suggests an orifice larger than Pinhole-L. This specific degradation likely highlights a vulnerability to the domain shift between the idealized synthetic training data and the noisier, phase-averaged experimental PIV inputs. Because the current snapshot-based architecture infers the orifice geometry purely from the instantaneous 2D input slice, the structural context of the narrow regurgitation jet (systole) does not carry over into diastole. Consequently, when driven solely by the low-magnitude approaching flow captured within the PIV window, the network appears to become more sensitive to the differences between PIV and the CFD training data and predicts an implausible flow field.

Similar behavior is observed in the Slot-L and Drop-XL cases. For Slot-L, the experimental recording resolves a complex double vortex ring, a topological feature entirely absent from the synthetic training distribution. As shown in Appendix~\ref{sec:appresults} (Fig.~\ref{fig:slotlphase},a), the $x$-$y$-plane reconstruction appears qualitatively reasonable, though the jet tapers unexpectedly near the right atrial wall. However, examining the $x$-$z$-plane (Fig.~\ref{fig:slotlphase},b) once again reveals a disjointed volumetric structure and a premature wall connection. The TTA step insufficiently corrects the baseline prediction outside the additionally supervised central area, leaving the outer fluid layers and their respective vortex structures temporally advanced compared to the corrected jet core.

The Drop-XL application further demonstrates both the model's capabilities and its susceptibility to out-of-distribution dynamics (Appendix Fig.~\ref{fig:dropxlphase}). During systole, the model achieves a coherent topology in the $x$-$y$-plane (Fig.~\ref{fig:dropxlphase},a), successfully extending the jet to the right atrial wall. However, there are disproportionately elevated velocity intensities parallel to this boundary, and temporal analysis reveals again a premature wall connection. The corresponding $x$-$z$-plane (Fig.~\ref{fig:dropxlphase},b) serves as another positive example, demonstrating a reasonable and coherent cross-sectional reconstruction during this fully connected systolic instant. In contrast, the onset of the diastolic phase (Fig.~\ref{fig:dropxlphase},c) suffers from significant reconstruction failures, characterized by a disjointed jet with gaps and speckled areas of higher velocity. As previously observed with the Pinhole-L configuration, the network generally struggles to maintain structural coherence during the diastolic phase when driven by real-world data. For the Drop-XL case, the overall sensitivity to real-world data is likely further exacerbated by a hardware limitation observed during the physical experiment: the cardiac pump reached its power limit, which created clipped input waveforms and thereby unfamiliar driving pressure dynamics falling far outside the training distribution (see Appendix~\ref{sec:validation} and Fig.~\ref{fig:planerender_dropxl},c).

\begin{figure*}
    \centering
    \includegraphics[width=\textwidth]{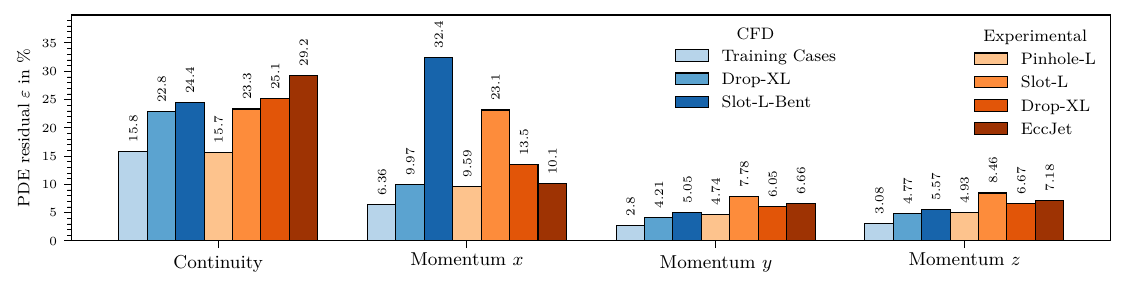}
    \vspace{-8mm}
    \caption{URANS residuals aggregated over the cardiac cycle. Blue bars denote results driven by synthetic CFD input, including the aggregate of all pre-training cases and the two fine-tuned validation cases. Orange bars represent the fine-tuned experimental cases driven by real-world PIV data.}
    \label{fig:pde_barplot}
\end{figure*}

Due to the lack of dense 3D ground truth, the physical consistency of the experimental reconstructions is evaluated using Unsteady Reynolds-Averaged Navier-Stokes (URANS) PDE residuals, aggregated over the cardiac cycle in Fig.~\ref{fig:pde_barplot}. These residuals quantify physical inconsistency by calculating spatiotemporal L1 integral norms of the non-dimensionalized mass and momentum balance (Definition in Appendix~\ref{app:PDERes}). Because this formulation relies on absolute unnormalized error magnitudes, the resulting metrics are inherently sensitive to the volume and intensity of high-velocity regions, where convective forces and spatial gradients are largest. Consequently, the lower aggregate residuals observed for the pre-training pool likely reflect, in part, the inclusion of smaller orifice geometries that generate less severe jets compared to the isolated validation cases. 

Despite this mathematical caveat, the metric provides a useful baseline for order-of-magnitude comparisons between the synthetic and experimental applications. When driven by temporally sparser, noisy, real-world PIV data, the aggregate residual magnitudes for the experimental cases generally align with those of the synthetic validation cases. The $x$-momentum error predictably remains the highest across all configurations, reflecting the dominance of the primary jet trajectory along this axis. Ultimately, these results indicate that applying the purely data-driven model to experimental data maintains a level of macroscopic physical consistency comparable to the synthetic out-of-distribution baseline, without catastrophic degradation.

The bar plot also includes the EccJet configuration, which serves as a rigorous stress test due to its highly eccentric jet angle that is completely absent from the training distribution. Prior to adaptation, the baseline data-driven model fails to resolve this novel trajectory, incorrectly predicting a straight jet impinging on the right atrial wall. Following TTA, the hybrid model successfully reorients the jet direction to match the eccentric angle observed in the PIV data. However, limitations remain: the jet extends slightly past the PIV window, but fails to reach the bottom wall of the atrium. Additionally, remnants of the training distribution persist, visible as falsely elevated velocities along the right atrial wall, and the left ventricular streamlines are degraded, appearing overly uniform.

\begin{figure*}
    \centering
    \includegraphics[width=\textwidth]{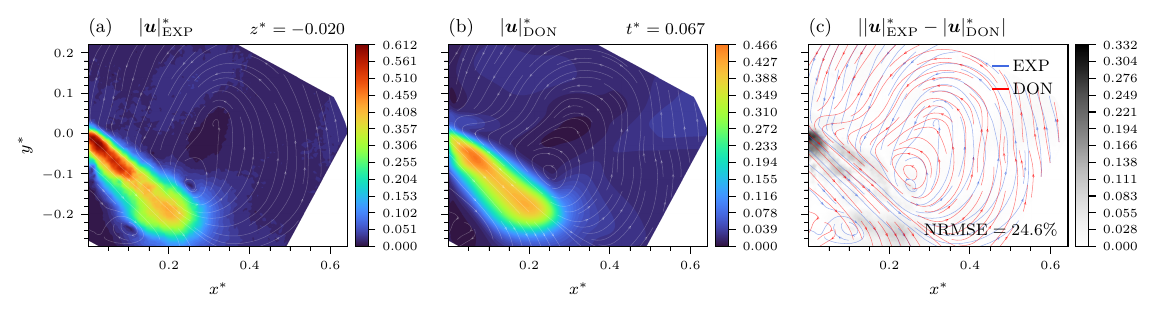}
    \vspace{-9mm}
    \caption{(a) Reference PIV velocity magnitude at $z^*=-0.02$ and $t^*=0.067$ for the EccJet orifice. (b) DeepONet prediction. (c) Absolute error magnitude with streamline comparison.}
    \label{fig:eccjet}
\end{figure*}

\begin{table}[]
\caption{\label{tab:eccjetvalues} NRMSE values for the velocity magnitude prediction compared against PIV measurements for the EccJet case.}
\begin{ruledtabular}
\begin{tabular}{cccc}
NRMSE in \% & $z^*=-0.02$ & $z^*=-0.04$ & $z^*=-0.06$ \\
\hline
$t^*=0.040$ & 28.2 & 62.0 & 96.3 \\
$t^*=0.053$ & 26.5 & 52.9 & 72.1 \\
$t^*=0.067$ & 24.6 & 43.1 & 52.6 \\
$t^*=0.080$ & 23.9 & 41.6 & 49.8 \\
$t^*=0.093$ & 21.0 & 35.1 & 43.9 \\
\end{tabular}
\end{ruledtabular}
\end{table}

Crucially, the EccJet dataset provides phase-resolved PIV measurements for adjacent $z$-planes, enabling a quantitative assessment of volumetric reconstruction. Fig.~\ref{fig:eccjet} compares the prediction against PIV data at a depth of $z=-2\,\mathrm{mm}$. Here, the model captures the general flow direction and large vortex structures, achieving an NRMSE of 24.6\%. However, Table~\ref{tab:eccjetvalues} demonstrates that reconstruction accuracy degrades sharply as the distance from the supervised center slice increases. At a deeper plane ($z=-6\,\mathrm{mm}$), the prediction exhibits a marked loss of definition, and the streamlines no longer agree to a satisfactory degree, causing the NRMSE to more than double to 52.6\%. This trend of increasing error and structural degradation is consistently observed across all other available time-steps. These findings confirm that this single-plane TTA strategy is currently insufficient to ensure reliable volumetric results across the full 3D domain.

%% file: sections/conclusion.tex
In this work, a hybrid Deep Operator Network (DeepONet) framework was proposed to reconstruct transient, three-dimensional hemodynamics from sparse 2D planar velocity measurements and phase-resolved boundary pressures. The results demonstrate that the framework can successfully infer plausible full-field 3D hemodynamics in a matter of minutes, bypassing the prohibitive computational costs associated with traditional transient CFD simulations. 
The two-step Test-Time Adaptation (TTA) strategy effectively corrects in-plane flow topologies even for out-of-distribution geometries, as demonstrated by the successful reorientation of the flow trajectory for the highly eccentric EccJet configuration.

Despite these advantages, the validation on synthetic data revealed critical limitations. Primarily, the TTA fine-tuning step is strongly localized to the supervised 2D measurement plane as it fails to propagate meaningfully through the domain. Consequently, whenever the pre-trained baseline prediction deviates significantly from the true 3D state, such as variations in the temporal onset of the jet, it leads to severe volumetric incoherence in the form of disjointed 3D flow structures at the boundary of the additionally supervised area. Similarly, the TTA is unable to correct errors stemming from geometric extrapolation anywhere but in the center plane of the left atrium (LA), which leaves the jet persistently narrow and can result in a severe underestimation of the integrated regurgitant volume.

When applied to experimental measurements, the framework broadly mirrors the capabilities and limitations observed with synthetic data, successfully capturing general flow topologies within the measurement plane and yielding integrated regurgitation volumes that align closely with physical reference measurements. However, this bulk agreement masks severe underlying limitations: the use of real-world data intensifies previously identified sources of error and introduces more erratic, non-physical artifacts, particularly speckled velocity structures in the unconstrained areas of the application case. This degradation is likely driven by the combined effects of inherent measurement noise, the domain shift between the idealized rigid orifice plate of the CFD training data and the flexible-boundary physical experiment, and the database’s inability to fully encompass the complex reality of the driving inputs. For instance, unexpected experimental constraints, such as the physical pump reaching its power limit, generated unfamiliar boundary waveforms that fell far outside the training distribution, further highlighting the purely data-driven model's susceptibility to out-of-distribution signals.

Future research should investigate anchoring the volumetric solution through (orthogonal) multi-plane PIV supervision, which would provide additional constraints to resolve the lateral propagation issues identified here. To improve temporal consistency, the integration of temporal learning structures, such as Temporal Convolutional Networks (TCNs), should be investigated as an alternative to the snapshot-based approach. Additionally, the framework's robustness could be enhanced by expanding the training database with higher-fidelity CFD data, implementing noise-based data augmentation to bridge the simulation-to-reality gap, and re-exploring the inclusion of physics-based losses. In summary, while the current approach provides a promising pathway for reduced-cost hemodynamic modeling, further development is required to achieve coherent 4D reconstructions.

%% file: sections/simdetails.tex
The internal fluid domain was discretized using the polyhedral and prism layer meshers within STAR-CCM+. The standard configuration utilized a global base cell size of $3.0\,\mathrm{mm}$, a single prism layer on wall surfaces, a cell growth rate of 1.2, and a maximum volumetric cell size of 200\%. Localized volumetric and surface refinements were implemented to balance spatial resolution with computational efficiency. Like the maximum cell size, these were defined relative to the base cell size. The comprehensive relative settings and resulting absolute cell sizes are detailed in Table \ref{tab:mesh_baseline}.

Regions subjected to low velocity gradients, such as the upper reservoir and the cardiac pump intake cylinder, were meshed coarsely to reduce the overall cell count. Conversely, steep flow gradients and complex geometric features required localized refinement. The narrow aortic outflow tracts and the primary regurgitant jet trajectory (``jet cylinder'') were refined using targeted volumetric controls. The ``aorta sphere'' listed in Table~\ref{tab:mesh_baseline} affected the converging shape of the LV leading up to the aorta. The highest cell density was featured in the immediate vicinity of the MROP opening (``orifice cylinder'') to improve the flow development accuracy at the sharp edges. As the physical dimensions of the orifice vary significantly between severity classes, a uniform value for both the volumetric and surface sizes was selected from the displayed range. For instance, the Slot range of MROPs used 13\% (Slot-S), 20\% (Slot-M) and 25\% (Slot-L).

\begin{table}
\caption{\label{tab:mesh_baseline} Mesh~parameters. The surface size settings simultaneously describe the prism layer thickness and the target and minimum surface sizes of the polyhedral mesher. The lower half of listed regions denote subsections found in the main (LV \& LA) domain part.}

\begin{ruledtabular}
\begin{tabular}{lrrrr}
 & \multicolumn{2}{c}{Volumetric Size} & \multicolumn{2}{c}{Surface Size} \\
Region & in \% & in mm & in \% & in mm \\
\hline
Main (LV \& LA)   & 100 & 3.00 & 50 & 1.50 \\
Reservoir     & 200 & 6.00 & 50 & 3.00 \\
Pump & 200 & 6.00 & 100 & 6.00 \\
Aorta         & 50  & 1.50 & 50 & 0.75 \\ \hline
Orifice Plate & - & - & 25 & 0.75 \\
Aorta Sphere  & 50  & 1.50 & - & - \\
Jet Cylinder  & 50  & 1.50 & - & - \\
Orifice Cylinder & 13-25 & 0.39-0.75 & 13-25 & 0.39-0.75 \\
\end{tabular}
\end{ruledtabular}
\end{table}

\begin{figure}
    \centering
    \includegraphics[width=1\linewidth]{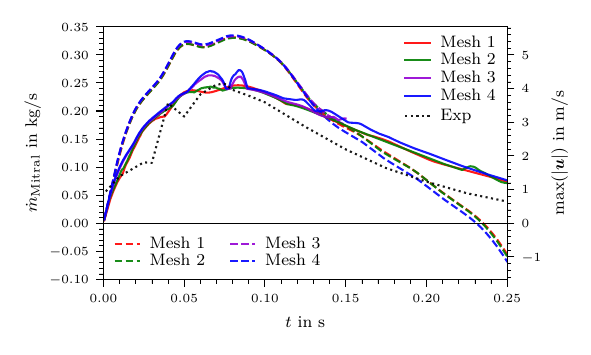}
    \vspace{-8mm}
    \caption{\label{fig:mesh_val_paper} Mass flow through Pinhole-L MROP (dashed) and maximum regurgitation jet velocity magnitude (solid) for the first systole of the 4 studied meshes. Experimental velocity values (black dotted) included from PIV.}
\end{figure}

Grid independence was investigated using the Pinhole-L geometry for four configurations of varying global base sizes as defined in Table~\ref{tab:meshes}. Compared were the maximum velocity magnitude and mass flow through the orifice for the first systole of the simulations from the cold start of zero-value initial conditions (Fig.~\ref{fig:mesh_val_paper}).

\begin{table}
\caption{\label{tab:meshes} Configurations evaluated during the mesh independence study for the Pinhole-L geometry. Relative refinement percentages are shown alongside their resulting absolute cell sizes.}
\begin{ruledtabular}
\begin{tabular}{lrrrrrr}
 & Base Size & \multicolumn{2}{c}{Aorta Sphere} & \multicolumn{2}{c}{Orifice Cylinder} & Cell \\
Mesh~& in mm & in \% & in mm & in \% & in mm & Count \\
\hline
1 & 5.0 & 50 & 2.500 & 25 & 1.250 & 145,678 \\
2\footnote{Selected baseline configuration. Other settings match with Table~\ref{tab:mesh_baseline}.} & 3.0 & 50 & 1.500 & 25 & 0.750 & 427,557 \\
3 & 2.0 & 75 & 1.500 & 25 & 0.500 & 1,042,883 \\
4 & 1.5 & 75 & 1.125 & 25 & 0.375 & 1,993,642 \\
\end{tabular}
\end{ruledtabular}
\end{table}

Throughout the simulations, the mass flow across the orifice remained very similar across all four grids. The difference of corresponding integrated values (RVol) remained within 1.5\%. Variations were observed in the peak velocity magnitude, where after $0.05\,\mathrm{s}$, the two finer and coarser meshes separate more clearly. Higher peak velocities are found in the tip of the regurgitant jet for the finer meshes, however the tip makes up only a small part of the overall structure. Consistent with the higher velocity, the jets reached slightly deeper into the LA at the same simulation time with increasing mesh density. While the flow for Mesh~3 converged more clearly towards the flow observed in the densest Mesh~4, Mesh~2 (427,557 cells) was ultimately selected for the multiple MROP dataset creation due to its lower computational cost. The development of the machine learning framework itself remained effective even without maximally accurate source data, and the training data can be refined at a later stage. Because the simulation data for Mesh~3 (1,042,883 cells) was already generated during this validation phase, it was used exclusively for the Pinhole-L data export.

Near-wall resolution was evaluated via the non-dimensional wall distance, $y^+ = (u^\star y) / \nu$. Here, $u^\star$ acts as a hybrid scale for the adaptive ``All-$y^+$'' wall modeling approach implemented in STAR-CCM+ and used in this work, which enables valid boundary conditions for a wide range of near-wall mesh densities.~\cite{siemens_digital_industries_software_simcenter_2024} In the viscous sublayer, $u^\star$ recovers the conventional friction velocity driven by wall shear stress ($u^\star \approx u_\tau = \sqrt{\tau_w/\rho}$). As the distance increases into the buffer and logarithmic regions, it is calculated non-iteratively by blending with a turbulent scale. In the Slot-L mesh (443,353 cells) as a representative case, the spatial maximum $y^+$ on the orifice plate wall fluctuated between 4 and 27 throughout the cardiac cycle. Across the remainder of the domain, the spatial maximum $y^+$ ranged between 11 and 61.

%% file: sections/validation.tex
\begin{figure*}
    \centering
    \includegraphics[width=1\linewidth]{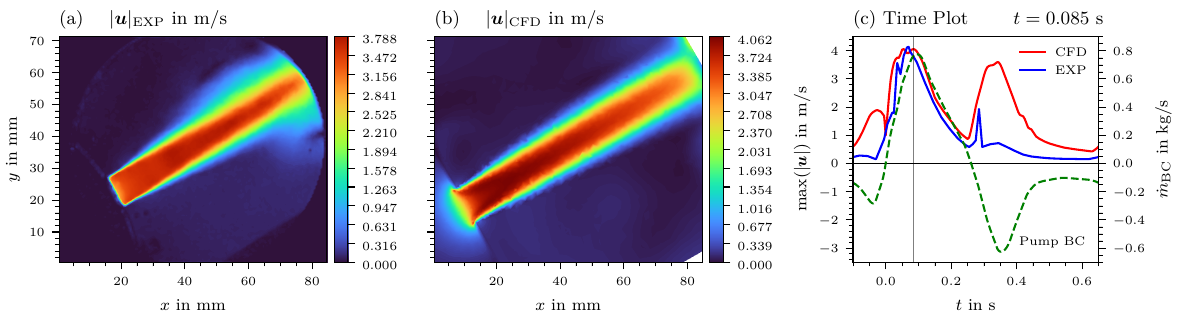}
    \vspace{-7mm}
    \caption{Pinhole-L velocity magnitude snapshot for (a) PIV and (b) CFD at $t=0.085\,\mathrm{s}$ and $z=0\,\mathrm{mm}$. Pump BC used in CFD and resulting maximum velocity magnitudes through the cardiac cycle shown in (c) with the vertical line marking the phase of (a, b).}
    \label{fig:planerender_pinholel}
\end{figure*}

This section presents a spatiotemporal comparison of the transient URANS simulations against physical measurements from the in-vitro setup. Because the datasets were synchronized manually, some phase misalignment is to be accounted for.

\begin{figure}[h!]
    \centering
    \includegraphics[width=1\linewidth]{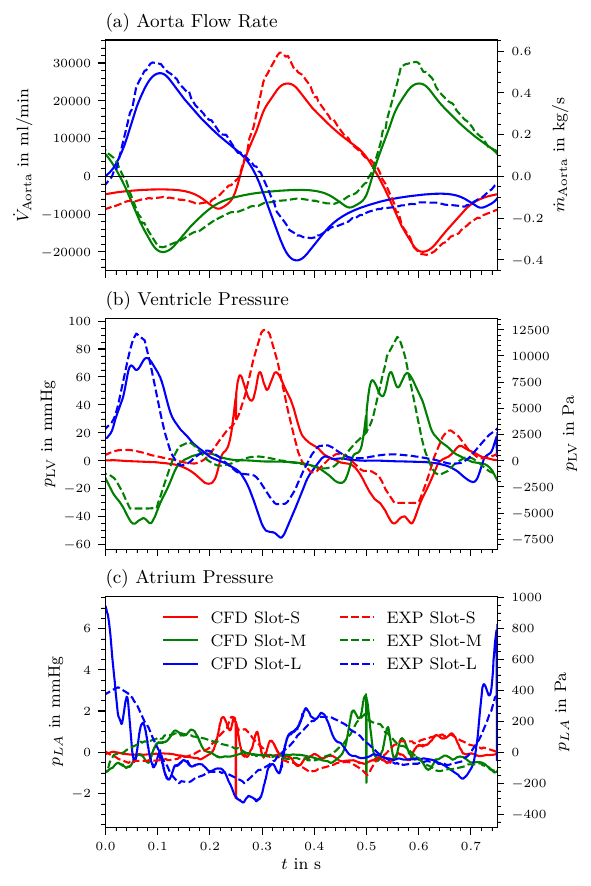}
    \vspace{-7mm}
    \caption{\label{fig:massflow_pressure_combined_paper} (a) Total aorta flow rate and (b) ventricle and (c) atrium pressure in the experiment (dashed) and CFD (solid). The three sizes of Slot MROPs are offset by $0.25\,\mathrm{s}$ for legibility. Two unit scalings shown each.}
\end{figure}

System-level behavior is evaluated against the flow rate ultrasonically measured through both aorta tubes and the pressure measurements at $\boldsymbol{x}_\mathrm{LV}$ and $\boldsymbol{x}_\mathrm{LA}$  across three Slot MROP sizes. The simulations satisfactorily resolve the general profile shape and amplitudes of the aortic flow rate (Fig.~\ref{fig:massflow_pressure_combined_paper},a), though the systolic peak is consistently underestimated. The pressure measurements in the left ventricle (LV) match the overall experimental trend (Fig.~\ref{fig:massflow_pressure_combined_paper},b), but the CFD model underpredicts the peak systolic and overpredicts the peak diastolic pressures. Furthermore, the experimental LV curves feature higher-frequency fluctuations that the numerical model smooths out. Conversely, in the left atrium (LA), the simulated pressure curves exhibit more high-frequency extrema than the physical experiment (Fig.~\ref{fig:massflow_pressure_combined_paper},c). The flexible $0.5\,\mathrm{mm}$ PVC foil from the physical setup can dynamically deform and might at least partially be responsible for mediating these pressure fluctuations, an effect absent in the rigid-walled CFD model. Overall, the MROPs appear to create less flow resistance in the simulation, as shown by lower aortic flow rate and lower peak LV pressures. This is corroborated by the overestimation of the overall regurgitation volume (RVol) discussed in the main text (Fig.~\ref{fig:rvol}).

\begin{figure*}
    \centering
    \includegraphics[width=1\linewidth]{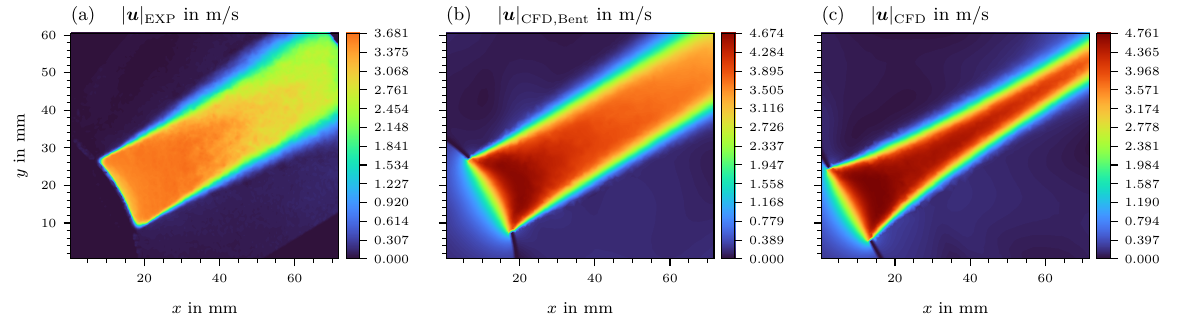}
    \vspace{-6mm}
    \caption{Comparison of (a) PIV and CFD velocity magnitude for Slot-L with (b) bent and (c) flat rigid orifice plate at $t=0.1\,\mathrm{s}$ and $z=0\,\mathrm{mm}.$}
    \label{fig:planerender_slotlphasebent}
\end{figure*}

\begin{figure*}
    \centering
    \includegraphics[width=1\linewidth]{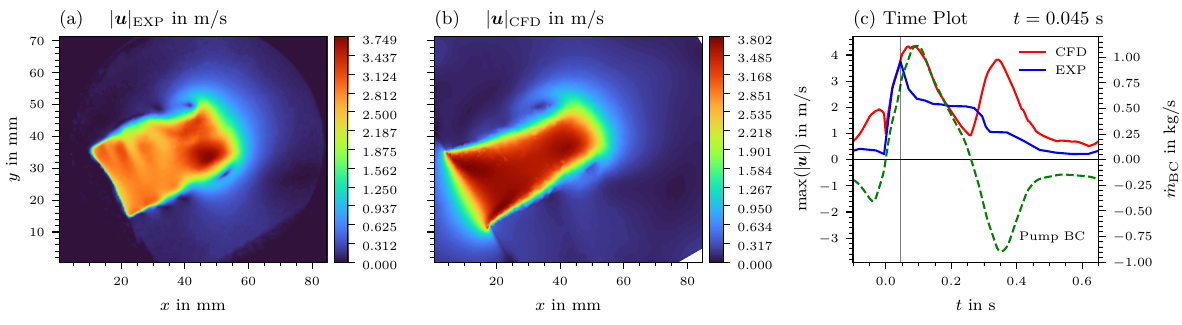}
    \vspace{-6mm}
    \caption{Drop-XL velocity magnitude snapshot for (a) PIV and (b) CFD at $t=0.045\,\mathrm{s}$ and $z=0\,\mathrm{mm}$. Pump BC used in CFD and resulting maximum velocity magnitudes through the cardiac cycle shown in (c) with the vertical line marking the phase of (a, b).}
    \label{fig:planerender_dropxl}
\end{figure*}

The transient development of the regurgitant jet was evaluated by comparing the CFD velocity fields against phase-resolved 2D2C PIV measurements. Overall, the URANS model successfully captures the topological development and spatial shape of the jets, although velocity magnitudes are generally overestimated and temporally rise earlier and decay later than in the physical setup. This is consistent with the previously identified lower flow resistance across the MROP. The closest match was observed for the Pinhole-L configuration (Fig.~\ref{fig:planerender_pinholel}), where velocity magnitudes and the core regurgitation jet shape align closely. The temporal trace of the maximum velocity magnitude diverges after $t \approx 0.25\,\mathrm{s}$ (Fig.~\ref{fig:planerender_pinholel},c) due to the diastolic back-filling flow that occurs outside the PIV camera's field of view in the LV, whereas the CFD data resolves the complete domain. 

The impact of the orifice plate shape is demonstrated with the Slot-L configuration. The standard flat CFD geometry inaccurately renders the spatial jet shape  (Fig.~\ref{fig:planerender_slotlphasebent},c); however, introducing a static spherical wall profile (Slot-L-Bent) that mimics the maximum systolic foil deformation yields a shape that matches the experimental PIV profile significantly better (Fig.~\ref{fig:planerender_slotlphasebent},b), despite the magnitude remaining overestimated. Specifically, the global maximum velocity reaches $4.99\,\mathrm{m/s}$ for CFD versus $4.07\,\mathrm{m/s}$ for PIV. In a temporal trace of the maximum velocity, (analogous to Fig.~\ref{fig:planerender_pinholel},c), the entire experimental curve for Slot-L sits consistently below the CFD prediction throughout systole. This magnitude mismatch varied across configurations and was most severe for the thick-walled EccJet MROP. For this eccentric case, the gap between the temporal traces is even wider, peaking at $5.39\,\mathrm{m/s}$ for CFD against $3.46\,\mathrm{m/s}$ for PIV. Nevertheless, the numerical model successfully captured the highly angled trajectory of the jet.

Finally, a hardware limitation was observed in the oversized Drop-XL case (Fig.~\ref{fig:planerender_dropxl}). During physical testing, the cardiac pump reached its maximum power limit, causing the piston displacement waveform to abruptly clip after $t = 0.045\,\mathrm{s}$. This hardware-induced plateau was not modeled in the CFD simulation, which utilized the standard boundary condition, explaining the divergence in the velocity profiles. 

Despite these quantifiable deviations in absolute magnitudes and idealized boundary conditions, the numerical model was deemed reasonably accurate for the purpose of developing the machine learning framework, which at this stage is not contingent upon the source data being maximally accurate.


%% file: sections/defs.tex
\subsubsection{\label{app:SSE}Sum of Squared Errors}

To drive the neural network's adaptation during training, PhysicsNeMo Sym defaults to an L2 objective function. The L2 loss computes the Sum of Squared Errors (SSE) between the model predictions $\hat{y}_k$ and the ground truth target data $y_k$ across $k$ discrete evaluation points, incorporating optional pointwise weights $\lambda_k$:
\begin{equation}
    \mathcal{L_\mathrm{L2}}
= \sum_{k} \lambda_k \left| \hat{y}_k - y_k \right|^{2}
\end{equation}
The absolute magnitude of the loss scales with the number of points $k$, which can subsequently influence the balance of the various loss components.

\subsubsection{\label{app:NRMSE}Normalized Root Mean Square Error}

Network predictions are evaluated against ground truth data using the Normalized Root Mean Square Error (NRMSE). By normalizing the RMSE against the standard deviation $\sigma$ of the target data, NRMSE measures the error relative to the variance of the dataset:
\begin{equation}
    \mathrm{NRMSE}
    = \frac{\mathrm{RMSE}}{\sigma}
    = \frac{\sqrt{\frac{1}{N} \sum_{k} \left( \hat{y}_k - y_k \right)^{2}}}
           {\sqrt{\frac{1}{N} \sum_{k} \left( y_k - \bar{y} \right)^{2}}}
\end{equation}
An NRMSE of 1.0 can be achieved by a trivial model that predicts the global average for all points, while values below 1.0 show further predictive power. This quantifies the unexplained variance, complementing the $R^2$ score which measures the explained variance:\cite{51}
\begin{equation}
    R^2= 1 - \mathrm{NRMSE}^2
\end{equation}
The aggregate NRMSEs (covering both space and time) are not calculated as the arithmetic mean of the instantaneous NRMSEs (computed for a single snapshot). Snapshots during quiescent flow phases can have unexpectedly high NRMSE despite low absolute errors due to a low standard deviation being the denominator. Instead, data points across the entire spatiotemporal sequence are pooled to compute a single, global RMSE, which is then normalized against the standard deviation of that complete dataset. For this, a third of the available snapshots (100) per case, and $5\%$ of points per snapshot ($\approx70,000$) were used. The subsequently defined PDE residuals were evaluated on these same points.

\subsubsection{\label{app:PDERes}PDE Residuals}

Adding to the characteristic quantities defined in Section~\ref{sec:methodology:ssec:architecture}, the fluid's density $\rho$ and viscosity $\nu=\mu/\rho$ are assumed to be constant, and therefore:
\begin{equation}
    \rho_c=\rho=1086.0\, \mathrm{ kg}/\mathrm{m}^3, \:\:\:\mu_c=\mu= 2.9961\cdot10^{-3}\,\mathrm{Pa}\,\mathrm{s}
\end{equation}
Based on these characteristic scales, the non-dimensional Reynolds, Strouhal, and Euler numbers are defined as:
\begin{equation}
    \mathrm{Re} = \frac{\rho_c \ u_c \ l_c}{\mu_c}, \quad \mathrm{St} = \frac{l_c}{u_c \ t_c}, \quad \mathrm{Eu} = \frac{p_c}{\rho_c \ u_c^{2}} 
\end{equation}

To account for turbulence within the (U)RANS framework, the governing equations incorporate a variable effective viscosity, $\mu_{\text{eff}}^*$. Following the Boussinesq hypothesis, the Reynolds stresses are modeled analogously to viscous stresses via a turbulent viscosity formulation.\cite{48} Consequently, we define the non-dimensional effective viscosity as the sum of the baseline fluid viscosity and the turbulent viscosity:
\begin{equation}
    \mu_{\text{eff}}^* = \mu^* + \mu_{\text{t}}^* = 1 + \frac{\nu_{t,c}}{\nu_c} \cdot \nu_t^*(x_i,t)
\end{equation}
Here, $\nu_t^*$ is the non-dimensional, spatiotemporal turbulent viscosity field directly constrained during training and output by the neural network. In laminar regions governed purely by the base fluid viscosity, or if the turbulence model is intentionally disabled, $\mu_{\text{eff}}^*$ simply reduces to 1. 

To evaluate the neural networks, the Navier-Stokes equations (NSE) are computed in their compressible form, because the network outputs do not inherently guarantee a divergence-free velocity field. Standard simplifications that eliminate the divergence terms from the momentum equations are not applied. Moving all terms to one side, the momentum residuals $r_{\mathrm{mom},i}$ and the continuity residual $r_{\mathrm{conti}}$ take the following form, based on the compressible form implemented in PhysicsNeMo Sym: \cite{44,45}


\begin{eqnarray}
        r_{\mathrm{mom},i} =&& \underbrace{\mathrm{St} \, \frac{\partial u_i^*}{\partial t^*}}_{\text{Transient } T_i} + \underbrace{u_j^* \frac{\partial u_i^*}{\partial x_j^*}}_{\text{Convection } C_i} + \underbrace{\mathrm{Eu} \, \frac{\partial p^*}{\partial x_i^*}}_{\text{Pressure }P_i} \nonumber \\
        && \underbrace{- \frac{1}{\mathrm{Re}} \frac{\partial}{\partial x_j^*} \left( \mu_{\text{eff}}^* \frac{\partial u_i^*}{\partial x_j^*} \right)}_{\text{Shear }S_i} \nonumber \\
        && \underbrace{- \frac{1}{\mathrm{Re}} \left[ \frac{\partial}{\partial x_i^*} \left( -\frac{2}{3} \mu_{\text{eff}}^*\ \Theta \right) + \mu_{\text{eff}}^* \frac{\partial \Theta}{\partial x_i^*} \right]}_{\text{Dilatational }D_i} \nonumber \\
        && \underbrace{- \frac{1}{\mathrm{Re}} \left( \frac{\partial \mu_{\text{eff}}^*}{\partial x_j^*} \frac{\partial u_j^*}{\partial x_i^*} \right)}_{\text{Coupling } V_i} \label{pdemom} \\ 
        \quad \text{with}  \quad && r_{\mathrm{conti}} = \frac{\partial u_j^*}{\partial x_j^*} =: \Theta \label{pdeconti}
\end{eqnarray}

The viscous contribution is decomposed into the shear ($S_i$), dilatational ($D_i$), and viscosity-gradient coupling ($V_i$) terms. The dilatational term $D_i$ vanishes if the network's predicted velocity field is divergence-free ($\Theta = 0$). Concurrently, the coupling term $V_i$ becomes large in flow regimes featuring steep gradients in the network-predicted turbulent viscosity field. Removing $D_i$ and restructuring $S_i$ and $V_i$ under the assumption of incompressibility leads to the following constant-density, variable-viscosity form of the NSE that were used for the physics-informed training tests mentioned in Section~\ref{sec:methodology:ssec:strategy}: 
\begin{eqnarray}
     0 =&& \mathrm{St} \, \frac{\partial u_i^*}{\partial t^*} + u_j^* \frac{\partial u_i^*}{\partial x_j^*} + \mathrm{Eu} \, \frac{\partial p^*}{\partial x_i^*} \nonumber \\
     &&- \frac{1}{\mathrm{Re}} \frac{\partial}{\partial x_j^*} \left[ \mu_{\text{eff}}^* \left( \frac{\partial u_i^*}{\partial x_j^*} + \frac{\partial u_j^*}{\partial x_i^*} \right) \right]\quad \\
     0 =&&\frac{\partial u_j^*}{\partial x_j^*}
\end{eqnarray}

Finally, to quantify PDE satisfaction of the models, the L1 integral norms of the residuals defined in Eq.~(\ref{pdemom}) and (\ref{pdeconti}) are calculated as follows. The instantaneous spatial error $\varepsilon_{\mathrm{space}}(t)$ approximates the spatial integral of the absolute residual over the domain volume $\Omega$ at a specific time t:
\begin{eqnarray}
    &&\varepsilon_{\mathrm{space}}(t)
    = \int_{\Omega} \left| r(t) \right| \, d\Omega
    \approx \sum_{k} w_{\mathrm{space}} \left| r_k(t) \right| \nonumber \\
    &&\text{with} \quad w_\mathrm{space}=\frac{|\Omega|^*}{N_\mathrm{points}}
\end{eqnarray}
The spatial weight $w_\mathrm{space}$ is defined as the non-dimensional volume of the sampled domain ($|\Omega|^*=1.5421$ for the cropped extents as defined in Section~\ref{sec:extraction}) divided by the number of sampled points. The total spatiotemporal error $\varepsilon_{\mathrm{total}}$ subsequently aggregates these spatial errors over the simulation duration using the temporal weight $w_\mathrm{time}$:
\begin{eqnarray}
    &&\varepsilon_{\mathrm{total}}
    = \int_{t} \int_{\Omega} \left| r(t) \right| \, d\Omega \, dt
    \approx \sum_{n} w_{\mathrm{time}} \, \varepsilon_{\mathrm{space}}(t_n) \nonumber \\
    &&\text{with} \quad w_\mathrm{time}=\frac{t_\mathrm{max}-t_\mathrm{min}}{N_\mathrm{snapshots}}
\end{eqnarray}
These weights ensure that the resulting error metric remains invariant to changes in spatial and temporal sampling density.

%% file: sections/appresults.tex
\begin{figure}[h]
  \centering

    \includegraphics[width=\linewidth]{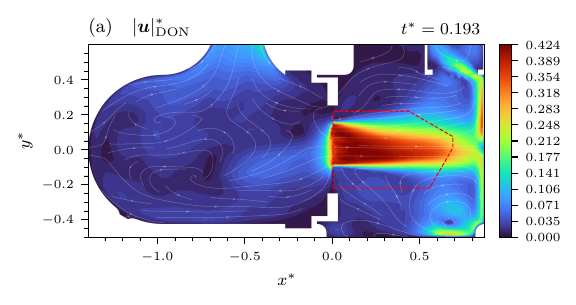}
    \vspace{-6.5mm}
      
    \includegraphics[width=\linewidth]{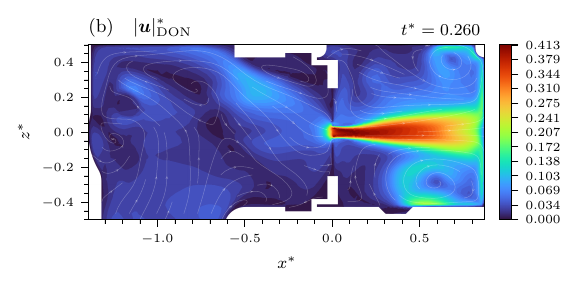}
    \vspace{-6.5mm}
    
    \includegraphics[width=\linewidth]{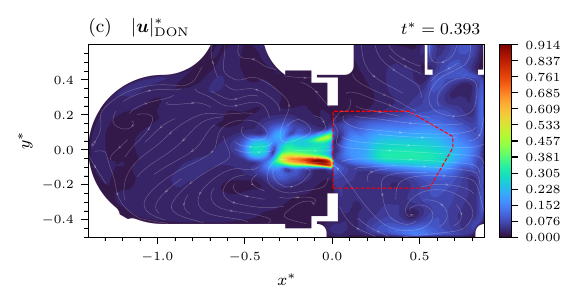}
    \vspace{-8mm}

  \caption{DeepONet velocity magnitude predictions for the Drop-XL orifice based on experimental PIV input. (a) $x$-$y$-plane at $z^* = 0$ and $t^* = 0.193$, (b) $x$-$z$-plane at $y^*=0$ and $t^* = 0.26$, (c) $x$-$y$-plane at $z^* = 0$ and $t^* = 0.393$.}
  \label{fig:dropxlphase}
\end{figure}

\begin{figure}[]
    \centering
    \includegraphics[width=1\linewidth]{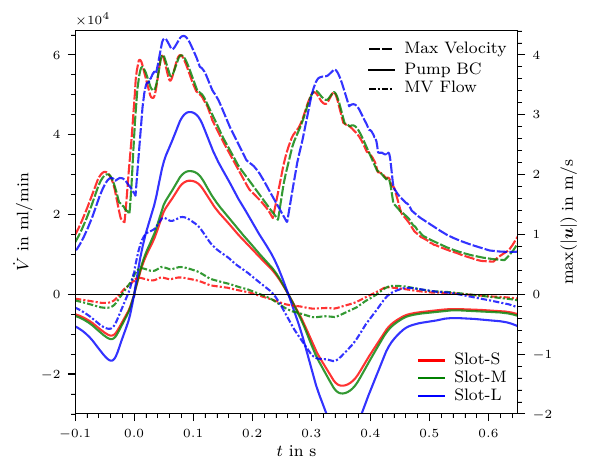}
    \vspace{-7mm}
    \caption{\label{fig:slotslm} Pump BC flow rate (solid), mitral valve flow rate (dash-dotted), and maximum velocity magnitude (dashed) across Slot-S (red), Slot-M (green), and Slot-L (blue) MROPs in the URANS CFD simulation.}
\end{figure}

\begin{figure}
  \centering

    \includegraphics[width=\linewidth]{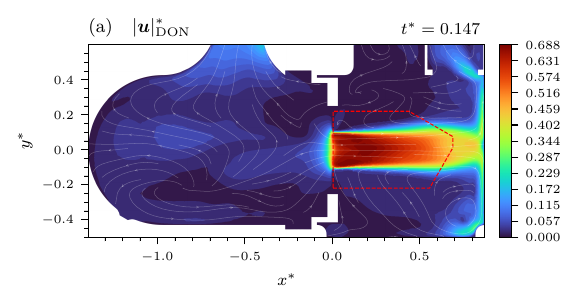}
    \vspace{-6.5mm}
    
    \includegraphics[width=\linewidth]{figures/don_larger_hybrid_dd2_pinholelphase_active_withp_pinholelphase_u_mag_y0.000_t0.086667_inference_paper.pdf}
    \vspace{-8mm}

  \caption{DeepONet velocity magnitude predictions for the Slot-L orifice based on experimental PIV input. (a) $x$-$y$-plane at $z^*=0$ and $t^*=0.147$, (b) $x$-$z$-plane at $y^*=0$ and $t^*=0.087$.}
  \label{fig:slotlphase}
\end{figure}

\begin{figure*}[] 
    \centering
    \includegraphics[width=\textwidth]{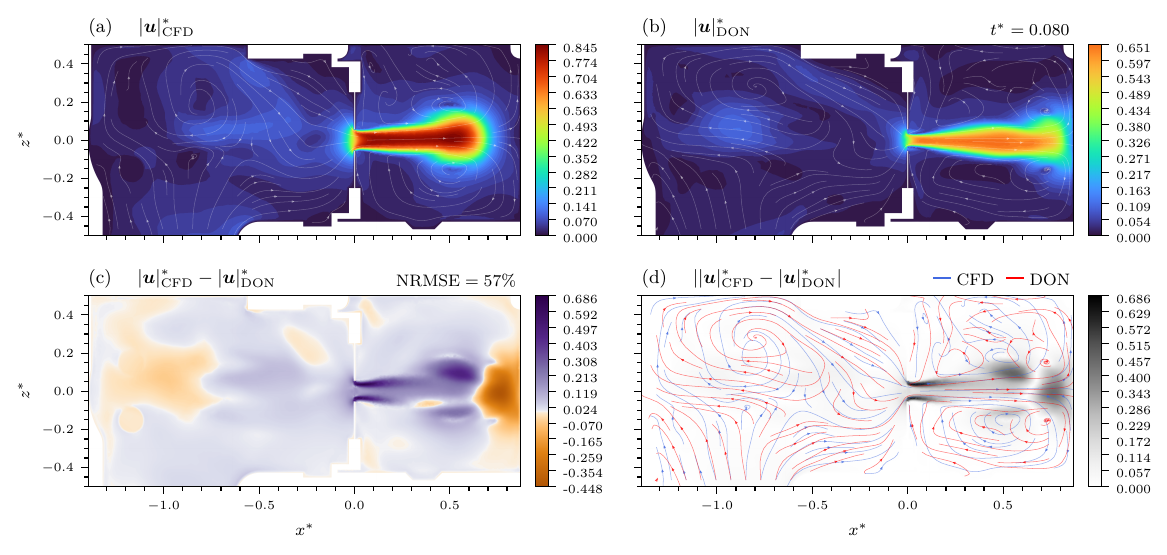}
    \vspace{-8mm}
    \caption{Velocity magnitude distribution at $y^*=0$ and $t^*=0.08$ for the Drop-XL orifice. (a) Reference CFD solution, (b) baseline DeepONet prediction before TTA. (c) Signed, non-linearly displayed and (d) absolute error magnitude with streamline comparison.}
    \label{fig:preTTA}
\end{figure*}

\begin{figure*}[] 
    \centering
\includegraphics[width=\textwidth]{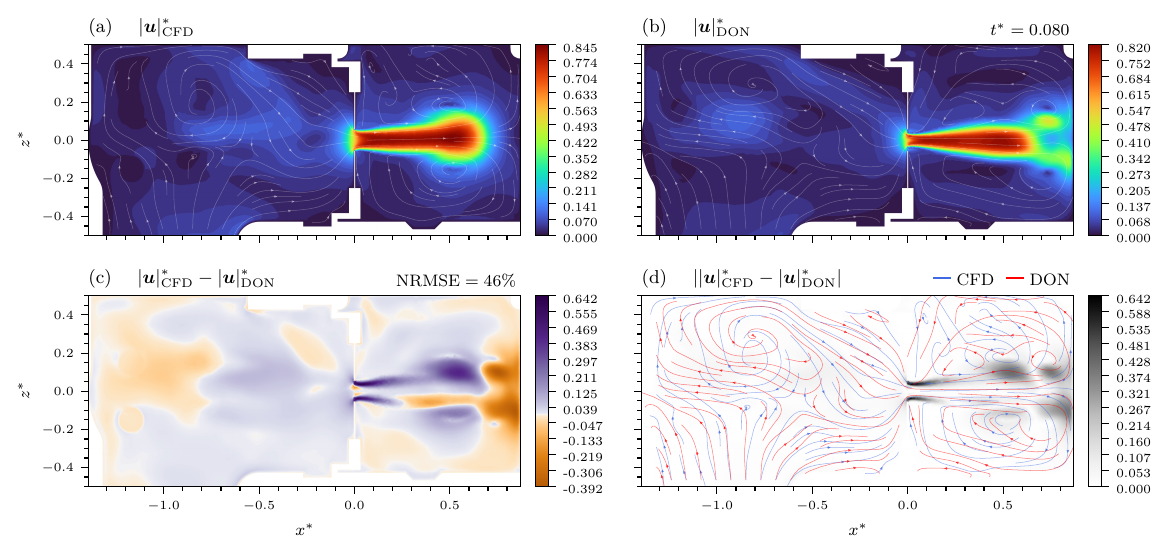}
    \vspace{-8mm}
    \caption{Velocity magnitude distribution at $y^*=0$ and $t^*=0.08$ for the Drop-XL orifice. (a) Reference CFD solution, (b) hybrid DeepONet prediction after TTA. (c) Signed, non-linearly displayed and (d) absolute error magnitude with streamline comparison.}
    \label{fig:postTTA}
\end{figure*}
